\documentclass[prd,amsmath,amssymb,longbibliography,12pt,nofootinbib]{revtex4-1}
\usepackage[normalem]{ulem}
\usepackage{soul}
\usepackage[utf8]{inputenc}
\usepackage{lineno,hyperref}
\usepackage{amsfonts,amssymb,amsmath}
\usepackage{epsfig}
\usepackage{mathrsfs}
\usepackage{graphicx}

\usepackage{xcolor}
\usepackage{appendix}

\begin{document}

\title{Thermal Krylov complexity and entropy of inflationary modes with different comoving-wavenumber dependences }
\author{Tao Li$^1$ and Lei-Hua Liu$^2$  }
\email{liuleihua8899@hotmail.com}
\affiliation{$^1$Department of Physics, Guizhou Minzu University, Guiyang 550025, China}
\affiliation{$^2$Department of Physics, College of Physics, Mechanical and Electrical Engineering, Jishou University, Jishou 416000, China }
\begin{abstract}

We study thermal Krylov complexity and Krylov entropy for primordial scalar modes in three inflationary models with different comoving-wavenumber dependences. These models are canonical single-field inflation, a non-trivial sound speed model, and a modified dispersion relation model. We construct a purified two-mode thermal state and first study its unitary Krylov evolution in an ideal closed system. In the parameter range considered here, the sound-speed correction increases the mean values and oscillation amplitudes of the Krylov quantities. The modified dispersion relation produces a stage-dependent oscillatory structure. We then introduce an independent effective open-system parametrization. The normalized Krylov distribution is controlled by a geometric ratio that combines thermal occupation, coherent spreading, and effective attenuation. We derive explicit expressions for Krylov complexity and Krylov entropy. We also obtain their low- and high-temperature limits. The sound-speed correction enhances the geometric ratio in the adopted parameter range. For the modified dispersion relation, coherent spreading and effective attenuation compete. These results show how thermal occupation and modified mode frequencies control Krylov-space spreading in the three inflationary models.

\end{abstract}
\maketitle
\section{Introduction}
\label{sec:introduction}

Quantum fluctuations generated during inflation provide the seeds of cosmic microwave background anisotropies and large-scale structure. Their enhancement on small scales may also be relevant to primordial black hole production. Quantum-information measures provide a complementary description of primordial state evolution. They can characterize state spreading, squeezing, and coherence. Circuit complexity measures the minimum cost of preparing a target state from a chosen reference state through a set of allowed unitary transformations \cite{Susskind:2014moa,Jefferson:2017sdb,Nielsen:2005mkt}. In holographic settings, circuit complexity has been conjectured to be related to the growth of black-hole interiors, motivating the complexity-volume and complexity-action proposals \cite{Hartman:2013qma,Susskind:2014moa,Susskind:2014rva}. In quantum field theory, Nielsen's geometric approach reformulates state preparation as an optimization problem on the manifold of unitary transformations, with the complexity determined by a minimal-cost path \cite{Jefferson:2017sdb}. Complexity methods have also been applied to black-hole information and spacetime evolution \cite{Zhao:2019nxk,Chandra:2021kdv}.

Krylov complexity provides a complementary description of dynamical state or operator growth \cite{Parker:2018yvk}. The Lanczos algorithm generates an orthonormal Krylov basis. Krylov complexity measures the mean position of the evolving probability distribution along the Krylov chain. Circuit complexity depends on a gate set and a cost function. Krylov complexity is instead constructed from the dynamics and the chosen initial state or operator. It has therefore become useful in studies of operator growth, scrambling, and quantum chaos \cite{Parker:2018yvk,Barbon:2019wsy}. Many recent works extend Krylov complexity to investigate early-universe quantum fluctuations \cite{Zhai:2025abc}. The resulting basis allows us to calculate Krylov complexity, Lanczos coefficients, and Krylov entropy \cite{He:2025guu,Zhai:2025abc,Jafarizadeh:2006woc}. Ref. \cite{Muck:2022xfc} presents practical evaluation schemes for orthogonal polynomial families. Krylov methods have been applied to many-body systems, conformal field theories, open quantum systems, and quantum field theories \cite{Parker:2018yvk,Rabinovici:2020ryf,Jian:2020qpp,He:2022ryk, Basu:2025ubf,Kar:2021nbm,Patramanis:2021lkx,Cao:2020zls,Trigueros:2021rwj,Heveling:2022hth,Dymarsky:2021bjq,Caputa:2021ori,Caputa:2022eye,Erdmenger:2023wjg,Patramanis:2023cwz,Hashimoto:2023swv,Vasli:2023syq,Domingo:2023kjr,Gill:2023umm,Bhattacharjee:2023uwx,Adhikari:2022whf,Ma:2026ksc,Liu:2026cxb}. These quantities provide related but non-equivalent information about spreading in the chosen Krylov basis. Refs. \cite{Camargo:2023eev, Bhattacharjee:2022vlt} studied Krylov operator complexity in quantum billiard systems, while Ref. \cite{Huh:2023jxt} uncovered universal spread complexity properties in saddle-dominated scrambling regimes.

Thermal effects can modify primordial fluctuations when the perturbation sector begins in a non-vacuum state or interacts with additional degrees of freedom, as in warm-inflation or open system descriptions. Thermal effects change the occupation numbers of field modes. They can also modify primordial correlation functions and power spectra \cite{Berera:1995ie}. If the perturbation sector is treated as an open system, interactions with unobserved degrees of freedom may generate dissipation and decoherence. Such interactions can modify the Lanczos coefficients and suppress coherent Krylov-space spreading \cite{Linardopoulos:2022wol,Li:2024ljz,Liu:2025caj,Zheng:2026vne,Liu:2026cxb,Ma:2026ksc,Liu:2026mzz}. During post-inflation preheating, parametric resonance generates abundant particles and drives thermalization; temperature governs chaotic intensity and separates weakly integrable low-temperature dynamics from strongly chaotic high-temperature behavior \cite{Maldacena:2015waa,Kofman:1994rk,Chakraborty:2023lpr}. Thermal Krylov dynamics in early-universe models remain less explored. Ref. \cite{Li:2024ljz} studied thermal Krylov complexity during inflation, radiation domination, and matter domination. Thermal contributions are physically relevant in models containing a radiation bath, non-vacuum initial conditions, or explicit system-environment couplings. They are model-dependent and must be introduced with clearly stated assumptions. Thermal effects in complexity growth were discussed in Ref. \cite{Kar:2021nbm}. This work first constructs a purified two-mode thermal state. The state evolves unitarily in an enlarged Hilbert space. We then introduce an independent effective open system description. The latter describes environmental damping \cite{Liu:2025caj,Li:2024kfm,Zhai:2024tkz,Zhai:2025abc,Bhattacharya:2022gbz}.

A non-trivial sound speed arises in k-inflation, DBI inflation, and effective single-field theories obtained after heavy degrees of freedom are integrated out \cite{Armendariz-Picon:1999hyi,Hikida:2004yc,Cai:2018tuh,Adhikari:2022oxr,Guth:1980zm}. It changes the \(k^2\) term in the mode frequency to \(c_s^2k^2\). It also shifts the sound-horizon crossing condition. A modified dispersion relation introduces higher-order dependence on the physical wavenumber. These terms become important when the physical wavenumber approaches the characteristic scale \(M\) \cite{Bojowald:2007cd,Kiritsis:2009sh,Naik:2001pr,Cai:2009hc}. Many existing studies of cosmological Krylov complexity focus on vacuum initial states. We compare the two modified models with canonical single-field inflation in the same two-mode thermal framework. The two corrections produce different Krylov-space evolution because they modify the mode frequency in different ways. The Lanczos coefficients characterize spreading in the chosen Krylov basis \cite{Li:2024kfm}. Previous studies examined cosmological complexity in non-trivial sound speed models and modified dispersion relation models \cite{Liu:2021nzx,Li:2021kfq,Liu:2026mzz,Li:2024kfm,Li:2023ekd}. Modified dispersion relations have also been considered in several ultraviolet and quantum-gravity-inspired scenarios \cite{Cai:2009hc,Armendariz-Picon:2003jjq,Armendariz-Picon:2006vgx,Magueijo:2008sx,Martin:2000xs,Arkani-Hamed:2003pdi,Bojowald:2006zb,Jacobson:2000gw,Cai:2007gs,Cai:2009in,Li:2009rt,Cai:2009zp,Cai:2012yf,Cai:2018tuh,Zheng:2017qfs,Chen:2017dhi,Bianco:2016yib,Pan:2015tza}. The present conclusions apply only to the mode functions and parameter ranges used in this work. 
 
The non-trivial sound speed and the modified dispersion relation produce different oscillatory structures in the Krylov quantities. These structures depend on the model parameters and the evolution stage. Krylov complexity and Krylov entropy provide complementary information about the three models \cite{Liu:2025caj,Li:2026jxx}. Higher-order momentum terms can generate short-time oscillations around the ultraviolet-to-infrared transition or horizon-crossing epoch. Within the models and parameters considered here, these oscillations differ from those produced by the sound speed modulation; a general proof of uniqueness is beyond the scope of the present analysis. This work provides a common thermal Krylov framework for comparing the three inflationary models. In the thermal construction, the effective temperature determines the occupation weights of the two-mode thermal state. The squeezing angle determines its phase structure. We also calculate the Lanczos coefficients, Krylov entropy, and effective damping coefficients. These quantities characterize evolution in the chosen Krylov representation.

The physical motivation for this study is that primordial perturbations may interact with other degrees of freedom. Short-wavelength modes, additional fields, reheating sectors, and gravitational degrees of freedom may act as an effective environment for the modes retained in the calculation. Tracing over these unobserved degrees of freedom produces a reduced state. The reduced evolution may contain decoherence and effective damping. In this work, we do not derive the reduced dynamics from a microscopic system-environment interaction. We use an effective parametrization of attenuation. The resulting Krylov complexity and Krylov entropy quantify how the selected primordial state spreads in the Krylov basis under coherent evolution and effective environmental damping. The open system construction used below should be understood as an effective parametrization rather than as a complete microscopic model of inflationary dissipation. In particular, the two-mode thermal state is a purification of a thermal density matrix and evolves unitarily in an enlarged Hilbert space. It is therefore distinct from a reduced-system description governed by a dissipative generator. The Krylov quantities are not direct cosmological observables. A quantitative observational connection requires separate calculations of the primordial correlation functions and related observables. We leave this analysis for future work.

We study canonical single-field inflation, a non-trivial sound speed model, and a modified dispersion relation model. Canonical single-field inflation is the reference case. Table \ref{tab:II} shows how the comoving-wavenumber enters the three models. We apply the same two-mode thermal construction to the three models. We calculate the effective mode temperature and squeezing angle. We then study thermal Krylov complexity and Krylov entropy in the ideal closed system and the effective open system. This comparison shows how the comoving-wavenumber dependence and environmental damping affect Krylov-space evolution.

This paper is organized as follows. Section \ref{sec:The thermal state} constructs the two-mode thermal state and separates thermal purification from effective attenuation. Section \ref{three cases} introduces the three inflationary models. Section \ref{krylov complexity and Krylov entropy in closed system} studies the ideal closed system. Sections \ref{krylov complexity and Krylov entropy in open system} and \ref{krylov complexity in open system} introduce the effective coefficients and the normalized open-system observables. Section \ref{summary} summarizes the results and discusses future work.

\section{The two-mode thermal state }\label{sec:The thermal state}

We begin with a bosonic mode of frequency $\omega$ at inverse temperature $\beta$. Its density matrix and a convenient purification are
\begin{align}
\hat{\rho}&=\frac{1}{z}\sum_{n=0}^{\infty}e^{-\beta E_n}|n\rangle\langle n|,
\label{eq:2.1}\\
|TFD\rangle&=\frac{1}{\sqrt z}\sum_{n=0}^{\infty}e^{-\beta E_n/2}|n\rangle\otimes|n\rangle_{\rm anc},
\label{eq:2.2}\\
|\Psi\rangle_{\phi}&=\frac{1}{\sqrt z}\sum_{n=0}^{\infty}(-1)^n e^{-2in\phi_k}e^{-\beta E_n/2}|n\rangle\otimes|n\rangle_{\rm anc}.
\label{eq:2.3}
\end{align}
Here $\beta=1/(k_{\rm B}T)$, $E_n=n\omega$, and $z=\sum_n e^{-\beta E_n}$. The ancillary Hilbert space is a copy of the original oscillator space. Tracing it out reproduces Eq.~\eqref{eq:2.1} \cite{Ali:2018fcz,Haque:2021kdm,Das:2024zuu}. The phased purification is also a two-mode squeezed state. The identification $\tanh r_k=e^{-\omega/(2k_{\rm B}T_k)}$ and $1/\cosh r_k=\sqrt{1-e^{-\omega/(k_{\rm B}T_k)}}$ gives
\begin{equation}
|\Psi\rangle_{\rm thermal}=\sqrt{1-e^{-\frac{\omega}{k_{\rm B}T_k}}}
\sum_{n=0}^{\infty}(-1)^n e^{-2in\phi_k}e^{-\frac{n\omega}{2k_{\rm B}T_k}}
|n\rangle_{\vec k}\otimes|n\rangle_{\rm anc}.
\label{eq:wave function of close syetem}
\end{equation}
Equation~\eqref{eq:wave function of close syetem} is a purification of the thermal density matrix. Its evolution in the enlarged Hilbert space is unitary. The temperature dependence must therefore be distinguished from irreversible attenuation. We first use this state to determine $T_k(\eta)$ and $\phi_k(\eta)$. We later introduce the independent effective coefficients $\mathcal A_k$ and $\mathcal B_k$. This separation distinguishes thermal occupation from the additional modification of Krylov-space spreading.

\section{Comoving-wavenumber dependence in the three inflationary models  }
\label{three cases}

We study three inflationary models. Canonical single-field inflation is the reference model. Its mode frequency contains the standard \(k^2\) term. The non-trivial sound speed model \cite{Armendariz-Picon:1999hyi,Garriga:1999vw} changes this term to \(c_s^2 k^2\). The sound speed resonance condition introduces the relation \(k=n k_\ast\). The modified dispersion relation introduces the function \(f(k_{\rm ph})\) \cite{Cai:2009hc}. The comoving-wavenumbers satisfy \(k=a k_{\rm ph}\). These three relations are summarized in Table \ref{tab:II}. We use the same background assumptions when we compare the three models. This setup allows us to study how different comoving-wavenumber dependences affect the two-mode thermal state and its Krylov-space evolution.

\subsection{Canonical single-field inflation}
\label{sec:canonical single-field inflation}

We decompose the inflationary system into a homogeneous background and linear perturbations. After combining the scalar-field and metric perturbations into the comoving curvature perturbation, the quadratic action can be expressed in terms of the Mukhanov-Sasaki variable. Throughout the numerical analysis we assume a quasi-de Sitter background with slowly varying $\epsilon $ and use conformal time derivatives denoted by primes.
\begin{equation}
S^{(2)}=\frac{1}{2}\int dt\,d^3x\,a^3\frac{\dot{\phi}_0^{,2}}{H^2}
\left[\dot{\mathcal R}^{,2}-\frac{1}{a^2}(\partial_i\mathcal R)^2\right],
\label{perturbated action}
\end{equation}
where \(H=\dot a/a\), \(\mathcal R=\psi+(H/\dot\phi_0)\delta\phi\), \(z=\sqrt{2\epsilon}\,a\), and \(\epsilon=-\dot H/H^2=1-\mathcal H'/\mathcal H^2\). We use the Friedmann--Lema\^itre--Robertson--Walker background metric
\begin{equation}
ds^{2} =a(\eta ) ^{2}\big(-d\eta ^{2}+d\vec{x} ^{2} \big),
\label{FRW metric}
\end{equation}
where \(\vec{x}=(x,y,z)\) is the three-dimensional spatial vector, and \(a(\eta)\) is the scale factor with respect to conformal time \(\eta\). Conformal time is more suitable for inflationary calculations, as it simplifies the description of horizon dynamics. To define the Mukhanov–Sasaki variable, we perturb the FRW metric in the following form
\begin{equation}
ds^2=a^2(\eta)\left[-\left(1+2\psi(\eta,\boldsymbol{x})\right)d\eta^2
+\left(1-2\psi(\eta,\boldsymbol{x})\right)d\boldsymbol{x}^2\right],
\label{perturbation of metric}
\end{equation}
where \(\psi(\eta,\boldsymbol{x})\) is the metric perturbation. We then rewrite the quadratic action \eqref{perturbated action} using the Mukhanov–Sasaki variable
\begin{equation}
  S^{(2)}=\frac{1}{2}\int d\eta d^{3}x\bigg[{v}'^{2}-(\partial _{i}v)^{2}+(\frac{{z}'}{z})^{2}v^{2}- 2{v}'  v\frac{{z}' }{z}\bigg] ,
  \label{eq:4.2}
\end{equation}
where \(v =z\mathcal{R}\). In order to construct the Hamiltonian operator, we need to define the field operator
\begin{equation}
\pi(\eta,\vec{x})=\frac{\delta L}{\delta{v}'(\eta,\vec{x})} ={v}'-\frac{{z}'}{z}v.
\end{equation}
We then construct the Hamiltonian via $H=\int d^3x (\pi v'-\mathcal{L} )$
\begin{equation}\label{hamilton of canonical single-field inflation}   
H= \frac{1}{2}\int   ~d^3x  \left [ {\pi}^2+(\partial _i v)^2+\frac{ {z}'}{z}(\pi v+v\pi) \right ].
\end{equation}
Using the Fourier decomposition
\begin{equation}\label{v}
\hat{v} (\eta,\vec{x})=\int \frac{d^3p}{(2\pi)^{3/2}} \sqrt{\frac{1}{2k}}\bigg(\hat{c }_{\text{anc} }^{\dagger}v_{\text{anc}}^{\ast }(\eta )e^{-i\vec{p\cdot }\vec{x}}+ {\hat{c}_{\text{anc}}}v_{\text{anc}}e^{i\vec{p\cdot }\vec{x}}\bigg),
\end{equation}
\begin{equation}\label{pi}
\hat{\pi} (\eta,\vec{x})=i\int \frac{d^3k}{(2\pi)^{3/2}}\sqrt{\frac{k}{2}}\bigg(\hat{c }_{\vec{k}}^{\dagger}u_{\vec k}^{\ast }(\eta )e^{-i\vec{k\cdot }\vec{x}}-\hat{c}_{\vec k}u_{\vec k}e^{i\vec{k\cdot }\vec{x}}\bigg).
\end{equation}
We therefore obtain the operator of Hamiltonian 
\begin{equation}
    \begin{split}
        \hat{H}_{k}= k\big(\hat c_{\text{anc}}^{\dagger }\hat c_{\text{anc}}+\hat c_{\vec{k}}\hat c^{\dagger }_{\vec{k}}\big)-i\frac{{z}'}{z}\big(\hat c_{\vec{k}}^{\dagger }\hat c_{\text{anc}}^{\dagger }-\hat c_{\vec{k}}\hat c_{\text{anc}}\big).
    \end{split}
     \label{eq:4.6}
\end{equation}
By using the Hamiltonian \eqref{eq:4.6} and wave function \ref{eq:wave function of close syetem}, we can obtain the evolution equations of $T_k(\eta)$ and $\phi_k(\eta)$ by solve Schrödinger equation
\begin{equation}\label{eq:5.3}
\begin{split}
   & {T}'_{k}=-\frac{4k_{\rm B}T_k^{2}}{\omega}\frac{a{}'}{a} \cos (2\phi_k)\sinh{\big(\frac{\omega}{2k_{\rm B}T_k}\big)}, \\&~~~{\phi}'_{k}=k-\frac{a{}'}{a}\sin (2\phi_k) \cosh{\big(\frac{\omega}{2k_{\rm B}T_k}\big)} ,
\end{split} 
\end{equation}
we assume a quasi-de Sitter background and treat \(\epsilon\) as approximately constant, thus we could obtain $\frac{z'}{z}=\frac{a'}{a}$ with $z=a\sqrt{2\epsilon}$. The corresponding evolutionary profiles can be obtained from the numerical solutions relating temperature and squeezing angle in Eq. \eqref{eq:5.3}.
\begin{figure}[h!]
    \centering
    \includegraphics[width=0.45\linewidth]{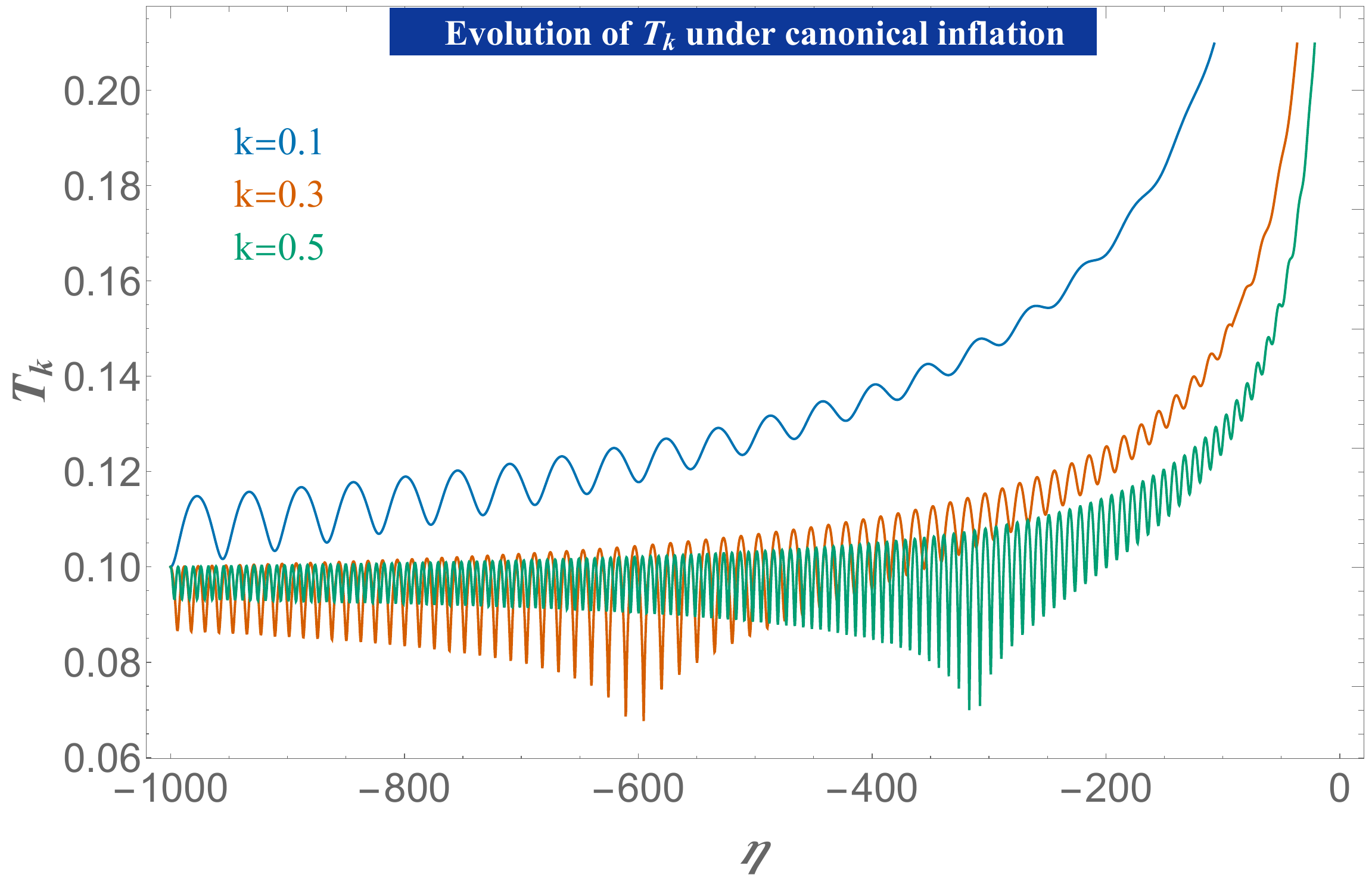}
    \includegraphics[width=0.45\linewidth]{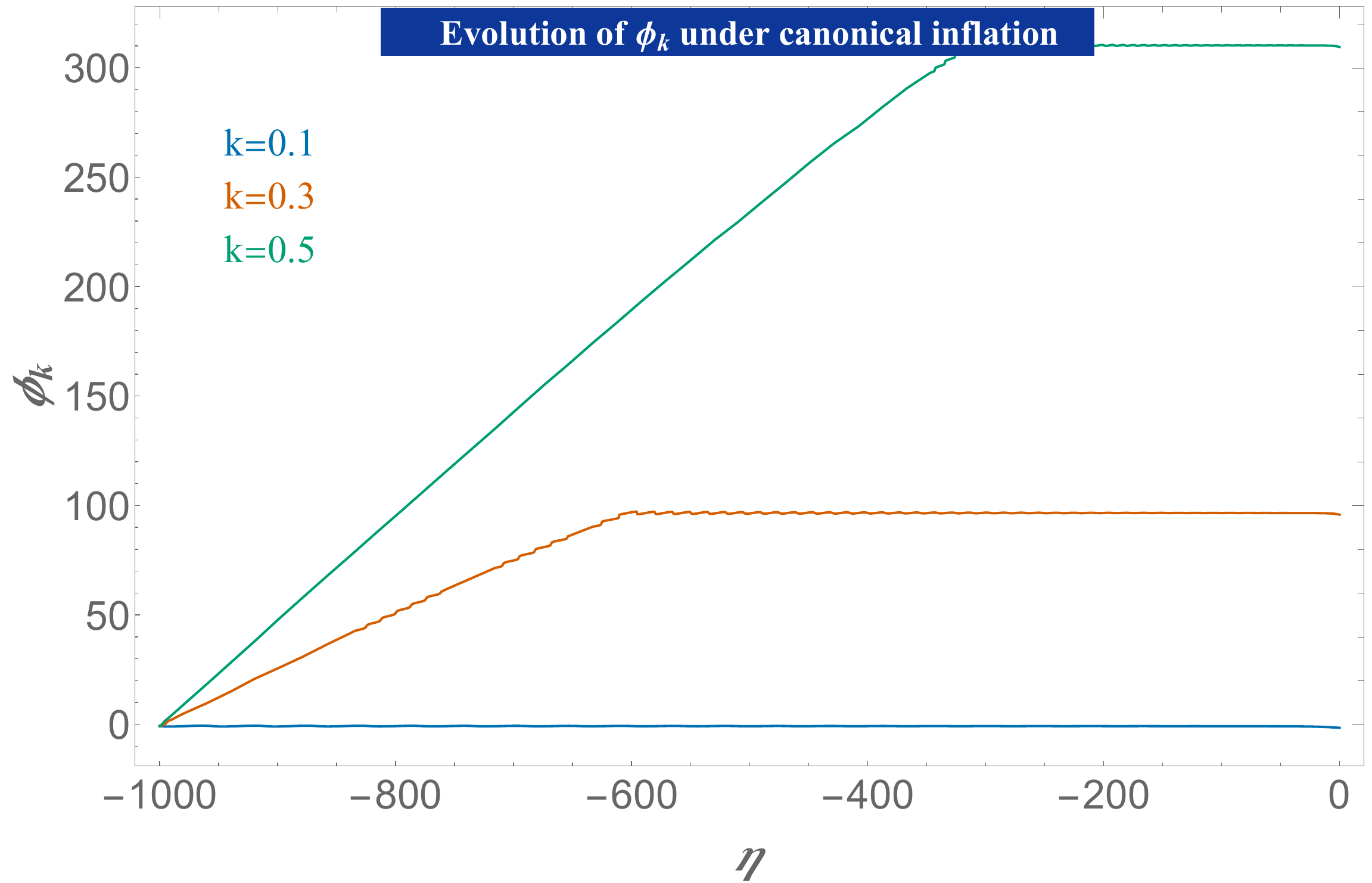}
    \caption{The effective mode temperature \(T_k(\eta)\) and squeezing angle \(\phi_k(\eta)\) as functions of conformal time \(\eta\) in canonical single-field inflation. The curves correspond to \(k=0.1,\ 0.3,\ 0.5\). We set \(\omega=1\) and \(k_{\rm B}=1\).}
    \label{fig:standard Tk and phik}
\end{figure}

Fig. \ref{fig:standard Tk and phik} shows the effective mode temperature $T_k(\eta)$ and squeezing angle $\phi_k(\eta)$ in canonical single-field inflation. The curves correspond to ($k=0.1, 0.3, 0.5$). The left panel shows $T_k(\eta)$. The curves exhibit oscillations over most of the plotted interval. Their mean values increase as $\eta$ approaches zero. Within the selected parameter range, the ($k=0.1$) mode has the largest effective temperature over most of the evolution. The differences between the curves become smaller near the end of the plotted interval. The right panel shows $\phi_k(\eta)$. The squeezing angle increases with conformal time and approaches a late-time plateau. Its final value depends on $k$. A larger value of $k$ gives a larger late-time squeezing angle for the initial conditions used here. The three values of $k$ correspond to different horizon-crossing times. The comparison is therefore made at fixed conformal time.

\subsection{Non-trivial sound speed}
\label{non-trivial sound speed}

A non-trivial scalar sound speed modifies the gradient term in the quadratic action. Such a sound speed can arise in k-inflation, DBI inflation \cite{Chen:2020uhe}, or effective single-field theories with integrated-out heavy modes; it is not necessarily a quantum-gravity correction by itself \cite{Armendariz-Picon:1999hyi,Garriga:1999vw}. Defining \(z = \frac{\sqrt{2\epsilon}\,a}{c_{s}}\), the corresponding equation of motion (EOM) reads
\begin{equation}
  v_k''  + \big(c_s^2k^{2}-\frac{z''}{z}\big) v_{k} = 0,
  \label{eom of sound speed}
\end{equation}
where we have performed the transformation to comoving-wavenumber space. One can verify that Eq. \eqref{eom of sound speed} smoothly reduces to canonical single-field inflationary equation when \(c_s^2=1\). From the effective field theory viewpoint, non-trivial sound speed can arise upon integrating out heavy modes \cite{Achucarro:2012sm}. In the present work, we focus on the sound speed resonance (SSR) mechanism \cite{Cai:2018tuh}, which admits a wide range of phenomenological realizations. The corresponding expression for \(c_s^2\) reads
\begin{equation}\label{Non-trivial sound speed with k}
    c_s^2=1-2\xi \big[1-\cos (k_{*}\eta)\big],
\end{equation}
Here, \(\xi\) is the amplitude of the sound speed modulation. The parameter \(k_\ast\) is the characteristic comoving-wavenumber. The resonance condition is written as \(k=n k_\ast\). Therefore, the non-trivial sound speed changes the mode frequency through both \(c_s(\eta)\) and its comoving-wavenumber dependence. Previous studies have shown that this resonance can enhance the scalar power spectrum on small scales. Such an enhancement may be relevant to primordial black hole production \cite{Cai:2018tuh}. From Eq. \eqref{eom of sound speed}, we readily derive the corresponding action
\begin{equation}
  S=\frac{1}{2}\int d\eta d^{3}x\bigg[{v}'^{2}-c_s^2(\partial _{i}v)^{2}+\big(\frac{{z}'}{z}\big)^{2}v^{2}- 2{v}'  v\frac{{z}' }{z}\bigg] .
  \label{hamiton of sound speed}
\end{equation}
Following the same procedure with Sec. \ref{sec:canonical single-field inflation}, we can obtain the Hamiltonian in terms of non-trivial sound speed
\begin{equation}
    H=\frac{1}{2}\int d^{3}x\bigg[\pi ^{2}+c_s^2(\partial_i v)^2+\frac{z'}{z}(\pi v+v\pi)\bigg].
    \label{eq:hf}
\end{equation}
Being armed with the Fourier decomposition \eqref{v} and \eqref{pi}, the Hamiltonian operator in momentum space can become as follows
\begin{equation}
\begin{split}
\hat H_k=\frac{k}{2}(c_s^2+1)\bigg(\hat c_{\text{anc}}^{\dagger}\hat c_{\text{anc}}+\hat c_{\vec k}\hat c_{\vec k}^{\dagger}\bigg)+\bigg(\frac{k}{2}(c_s^2-1)+i\frac{z'}{z}\bigg)\hat c_{\vec k}^{\dagger}\hat c_{\text{anc}}^{\dagger}+\bigg(\frac{k}{2}(c_s^2-1)-i\frac{z'}{z}\bigg)\hat c_{\vec k}\hat c_{\text{anc}}.
\label{hamilton of sound speed1}
\end{split}
\end{equation}
Following the same procedure, we obtain the $T_k(\eta)$ and $ \phi_k(\eta)$ of non-trivial sound speed as follows
\begin{equation}
  {T}'_{k}=-\frac{4k_{\rm B}T_k^{2}}{\omega}\sinh{\big(\frac{\omega}{2k_{\rm B}T_k}\big)}\bigg[\frac{a{}'}{a}\cos{(2\phi_k)}-\frac{k}{2}(c_s^2-1)\sin{(2\phi_k)}\bigg], 
  \label{eq:5.6}
\end{equation}
\begin{equation}
    \begin{split}
      {\phi}'_{k}=\frac{k}{2}(c_s^2+1)-\frac{k}{2}(c_s^2-1)\cos (2\phi_k)\cosh{\big(\frac{\omega}{2k_{\rm B}T_k}\big)} -\frac{a{}'}{a}\sin (2\phi_k)\cosh{\big(\frac{\omega}{2k_{\rm B}T_k}\big)}.
    \end{split}
    \label{eq:5.7}
\end{equation}
Subsequently, the corresponding evolutionary profiles can be derived from Eq. \eqref{eq:5.6}.
\begin{figure}[htbp]
    \centering
    \includegraphics[width=0.45\linewidth]{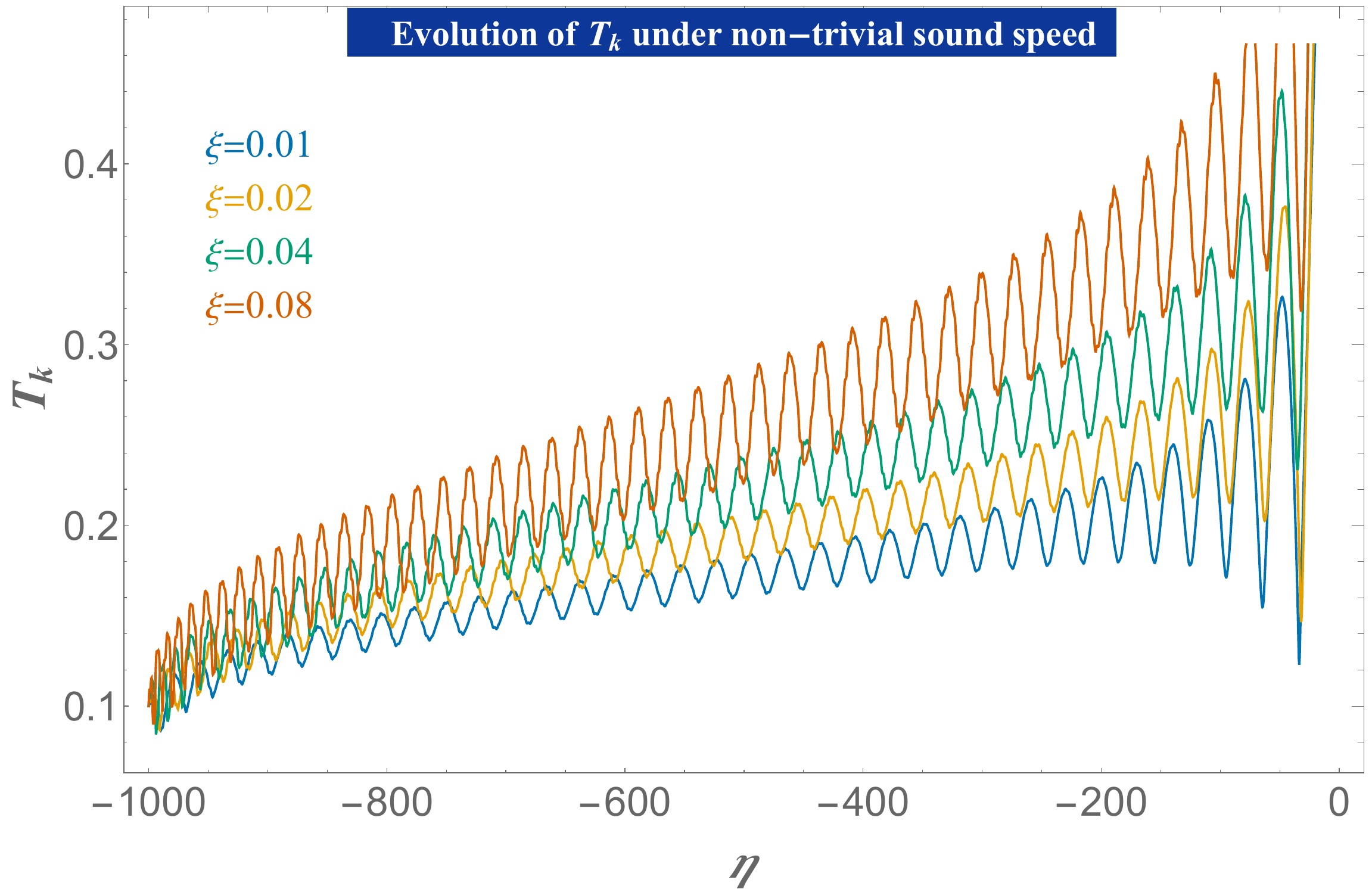}
    \includegraphics[width=0.45\linewidth]{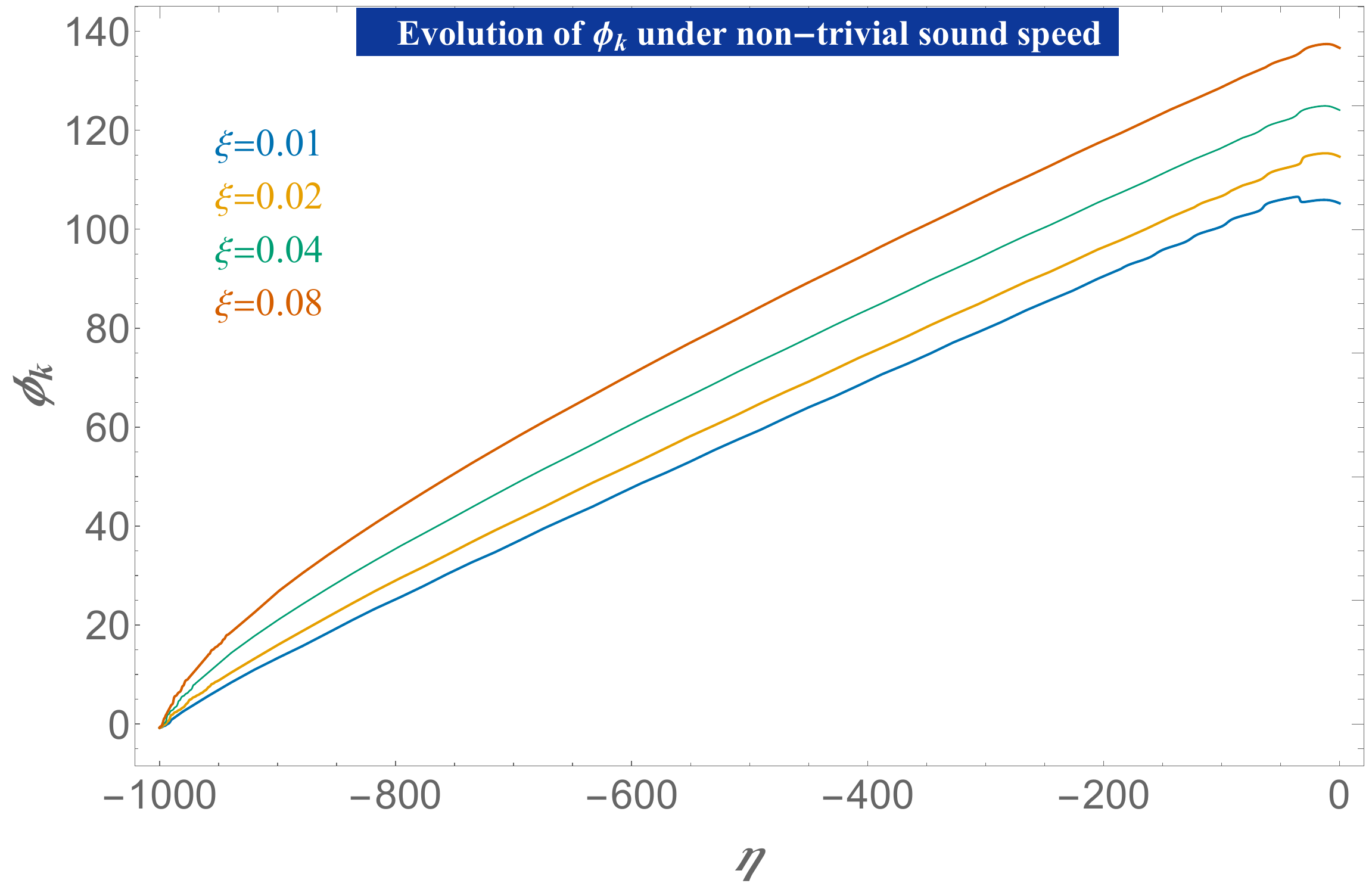}
    \caption{The effective temperature \(T_k(\eta)\) and squeezing angle \(\phi_k(\eta)\) as functions of conformal time \(\eta\) in the non-trivial sound speed model. The curves correspond to \(\xi=0.01,\ 0.02,\ 0.04,\ 0.08\). We set \(k=0.1\), \(\omega=1\), and \(k_{\rm B}=1\).}
    \label{fig:standard Tk and phik with Cs}
\end{figure}

Fig. \ref{fig:standard Tk and phik with Cs} shows the effective mode temperature $T_k(\eta)$ and squeezing angle $\phi_k(\eta)$ in the non-trivial sound speed model. We fix $k=0.1$. The parameter $\xi$ controls the amplitude of the sound speed modulation. The left panel shows $T_k(\eta)$. All curves exhibit oscillations. Their mean values and amplitudes increase during the evolution. Within the selected parameter range, a larger value of $\xi$ generally gives a larger effective temperature and a larger oscillation amplitude. The right panel shows $\phi_k(\eta)$. The squeezing angle increases with conformal time. Its growth becomes slower at late times. A larger value of $\xi$ generally gives a larger late-time value. The squeezing angle does not show the strong high-frequency oscillations found in $T_k(\eta)$. The oscillating sound speed modulates the instantaneous mode frequency. This modulation changes the effective mode temperature and the accumulated phase of the two-mode thermal state.

\subsection{The modified dispersion relation}
\label{sec:modified dispersion relation}

We next consider an effective modified dispersion relation in which the mode frequency departs from its relativistic form above a characteristic physical scale M \cite{Cai:2009hc}. Such a parametrization is motivated by several ultraviolet and Lorentz-violating scenarios, but the analysis below does not depend on assigning it to one unique microscopic theory. The corresponding action is 
\begin{equation}
\begin{split}
S= \frac{1}{2}\int d\eta d^3x\left [ {v}'^2-f^2(\partial _i v)^2+\big(\frac{z'}{z}\big)^2v^2-2\frac{ {z}'}{z}{v}'v  \right ],
\label{action of vi}
\end{split}
\end{equation}
with 
\begin{equation}
\begin{cases}
  & \text{ if }(k_{\rm ph})>M; f=(\frac{k_{\rm ph}}{M})^{\alpha }  \\
  & \text{ if }(k_{\rm ph})\le M; f=1
\end{cases},
\label{modified dispersion relation}
\end{equation}
where, \(k_{\rm ph}=k/a\) is the physical wavenumber. Equivalently, the comoving-wavenumber is \(k=a k_{\rm ph}\).

\begin{table}[h!]
\renewcommand\arraystretch{0.9}
\centering
\caption{Values of \(\alpha\) and their corresponding regimes in the adopted modified-dispersion parametrization. The case \(\alpha=0\) gives the standard dispersion relation. It corresponds to canonical single-field inflation in the present parametrization \cite{Guth:1980zm}.}\label{tab:I}
\centering
\begin{tabular}{lcccr}
\hline
Value of $\alpha$~~~~~~&~~~~~~$\alpha=0$~~~~~~&~~~~~~$0<\alpha<2$~~~~~~&$\alpha=2$ &~~~~~~~$\alpha>2$\\ \hline 
Regime&$IR$&$UV$& Ho\v{r}ava--Lifshitz cosmology \cite{Kiritsis:2009sh,Calcagni:2009ar}&$UV$\\
\hline
\end{tabular}
\end{table}
Starting from Eq. \eqref{action of vi}, we obtain the Hamiltonian 
\begin{equation}
\begin{split}
\hat H_k=\frac{k}{2}(f^2+1)\bigg(\hat c_{\text{anc}}^{\dagger}\hat c_{\text{anc}}+\hat c_{\vec k}\hat c_{\vec k}^{\dagger}\bigg)+\bigg(\frac{k}{2}(f^2-1)+i\frac{z'}{z}\bigg)\hat c_{\vec k}^{\dagger}\hat c_{\text{anc}}^{\dagger}+\bigg(\frac{k}{2}(f^2-1)-i\frac{z'}{z}\bigg)\hat c_{\vec k}\hat c_{\text{anc}}. 
\end{split}
\label{hamilton of dispersion relation}
\end{equation}
For the modified dispersion relation, we obtain
\begin{equation}
     {T}'_{k}=-\frac{4k_{\rm B}T_k^{2}}{\omega}\sinh{\frac{\omega}{2k_{\rm B}T_k}}\bigg[\frac{a{}'}{a}\cos{(2\phi_k)}-\frac{k}{2}(f^2-1)\sin{(2\phi_k)}\bigg], 
     \label{eq:5.61}
\end{equation}
\begin{equation}
    \begin{split}
     {\phi}'_{k}=\frac{k}{2}(f^2+1)-\frac{k}{2}(f^2-1)\cos (2\phi_k)\cosh{\big(\frac{\omega}{2k_{\rm B}T_k}\big)} -\frac{a{}'}{a}\sin (2\phi_k)\cosh{\big(\frac{\omega}{2k_{\rm B}T_k}\big)}.
    \end{split}
    \label{eq:5.71}
\end{equation}
Equations~\eqref{eq:5.61} and \eqref{eq:5.71} have the same structure as Eqs.~\eqref{eq:5.6} and \eqref{eq:5.7}. The replacement $c_s^2\rightarrow f^2$ distinguishes the two models. We obtain the corresponding profiles from the numerical solution of Eq.~\eqref{eq:5.71}.
\begin{figure}[htbp]
    \centering
    \includegraphics[width=0.45\linewidth]{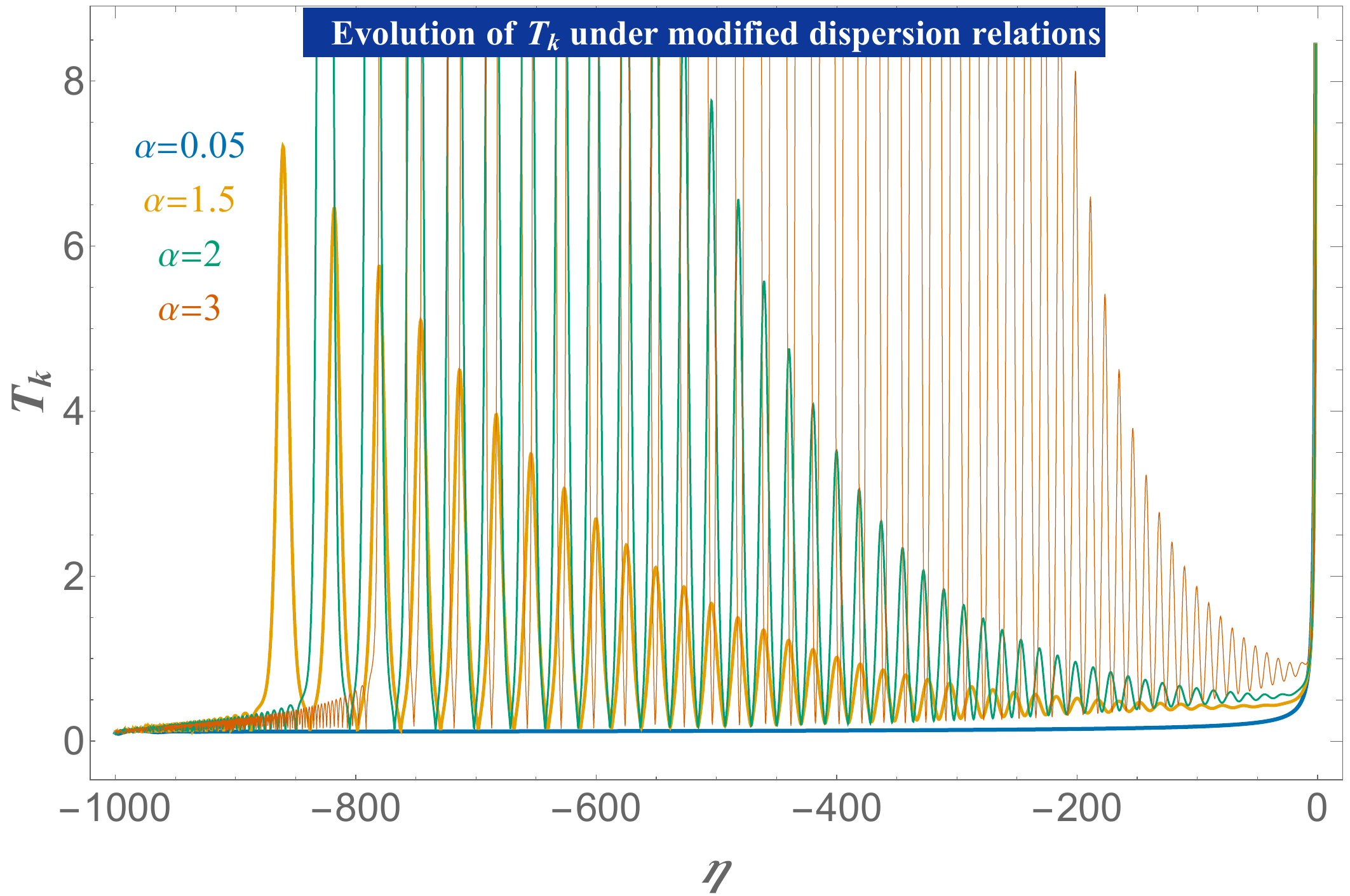}
      \includegraphics[width=0.45\linewidth]{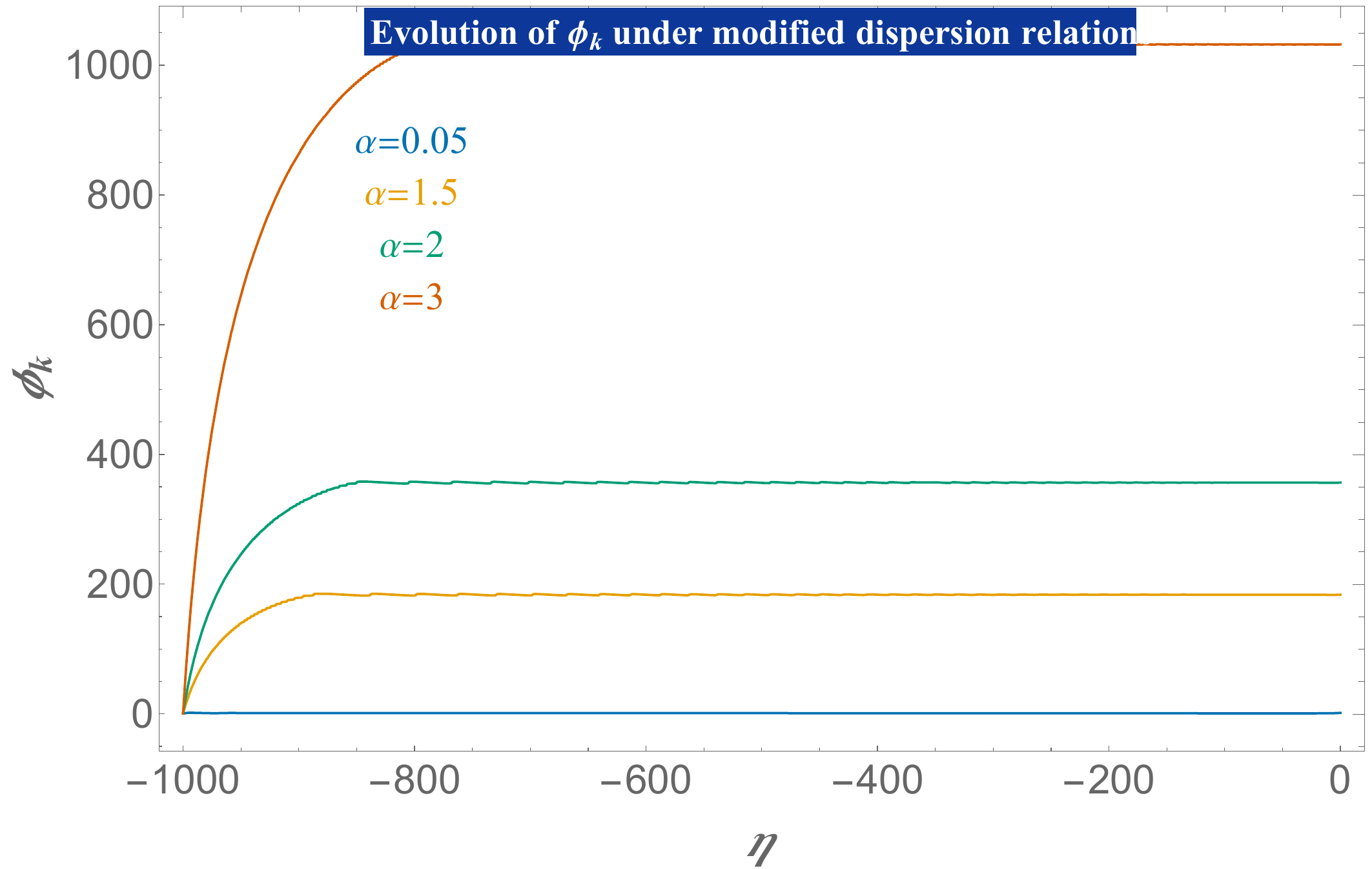}
    \caption{The effective mode temperature \(T_k(\eta)\) and squeezing angle \(\phi_k(\eta)\) as functions of conformal time \(\eta\) in the modified dispersion relation model. The curves correspond to $\alpha=0.01,1.5,2.0,3.0$. We set $k=0.1$, $\omega=1$, $k_{\rm B}=1$, $M=1$, and $k_{\rm ph}/M=1.5$.}
    \label{fig:standard Tk and phik with f}
\end{figure}

Fig.~\ref{fig:standard Tk and phik with f} shows the effective mode temperature $T_k(\eta)$ and squeezing angle $\phi_k(\eta)$ in the modified dispersion relation model. The left panel shows $T_k(\eta)$. The curves exhibit high-frequency oscillations over a large part of the plotted interval. Their amplitudes and mean values depend on $\alpha$. The effective temperature grows rapidly as $\eta$ approaches zero. Within the parameter range shown, larger values of $\alpha$ generally produce larger oscillation amplitudes and larger late-time values. The right panel shows $\phi_k(\eta)$. The squeezing angle increases rapidly during the early stage and then approaches a plateau. The plateau value depends on $\alpha$. Within the plotted parameter range, a larger value of $\alpha$ gives a larger late-time squeezing angle. These results show that the modified dispersion relation changes both the thermal occupation and the phase structure of the two-mode thermal state.

We further clarify the correspondence between the three inflationary models and comoving-wavenumber in Table \ref{tab:II}.

  \begin{table}[h!]
\renewcommand\arraystretch{0.9}
\centering
\caption{The correspondence between the three inflationary models and comoving-wavenumber.}\label{tab:II}
\centering
\begin{tabular}{lcl}
\hline
Model & ~~~~~~Mode-frequency term~~~~~~& Wavenumber relation \\ \hline 

Canonical single-field inflation & \(k^2\)& \(k\) is the comoving-wavenumber\\ \hline 
Non-trivial sound speed~\cite{Armendariz-Picon:1999hyi,Garriga:1999vw} & \(c_s^2 k^2\)& \(k=nk_\ast\)~~\eqref{Non-trivial sound speed with k}\\ \hline 
Modified dispersion relation~\cite{Cai:2009hc}& \(k^2 f^2(k_{\rm ph})\)& \(k_{\rm ph}=k/a\), or \(k=a k_{\rm ph}\)~\eqref{modified dispersion relation}\\ \hline 
\end{tabular}
\end{table}

Table \ref{tab:II} summarizes how the comoving-wavenumber enters the three models. The canonical model contains the standard \(k^2\) term. The non-trivial sound speed changes its coefficient and introduces the resonance relation \(k=nk_\ast\). The modified dispersion relation introduces a function of the physical wavenumber \(k_{\rm ph}=k/a\).

Sec. \ref{three cases} compares the effective mode temperature and squeezing angle in the three inflationary models. Canonical single-field inflation provides the reference evolution. The comoving-wavenumber $k$ changes both \(T_k(\eta)\) and \(\phi_k(\eta)\). In the non-trivial sound speed model, the sound speed changes the coefficient of the $k^2$ term. The resonance relation is $k=nk_\ast$. The periodic sound speed produces oscillations in the effective mode temperature. It also changes the late-time squeezing angle. In the modified dispersion relation model, the function $f(k_{\rm ph})$ changes the mode frequency. Its contribution becomes important when $k_{\rm ph}$ approaches the scale $M.$ The parameter $\alpha$ changes the oscillation pattern of the effective temperature and the late-time value of the squeezing angle.

 \section{Thermal Krylov complexity and Krylov entropy in the ideal closed system} 
 \label{krylov complexity and Krylov entropy in closed system}
 
In this section, we study thermal Krylov complexity and Krylov entropy in the ideal closed system. The two-mode thermal state evolves unitarily in the enlarged Hilbert space. No environmental damping is included in this section. In light of Eq. \eqref{eq:wave function of close syetem}, we obtain the corresponding operator wave function as follows
\begin{equation}\label{operator wave function of close syetem}
\psi_{n}= \sqrt{1-e^{-\frac{\omega}{k_{\rm B}T_k}}}(-1)^{n}e^{-2in\phi_k}e^{-\frac{n\omega}{2k_{\rm B}T_k}}.
    \end{equation}
The Lanczos recursion generates a Krylov basis from the chosen initial state or operator. Krylov complexity is the first moment of this distribution. Krylov entropy measures its spread \cite{Parker:2018yvk,Jafarizadeh:2006woc}. We state the definitions explicitly because their normalization is essential for comparing the three cosmological models. Also according to the definition of Krylov complexity, we could explicitly obtain
\begin{equation}
    \begin{split}
        K=\sum_{n=0}^{\infty}n|\psi_{n}|^{2}=\frac{e^{-\frac{\omega}{k_{\rm B}T_k}}}{1-e^{-\frac{\omega}{k_{\rm B}T_k}}}=\frac{1}{e^{\frac{\omega}{k_{\rm B}T_k}}-1}.
    \end{split}
    \label{Krylov complexity of close system}
\end{equation}
The Krylov complexity is influenced by $\omega$ and $T_k$ together, which is unlike the Krylov complexity without thermal effects, as shown in \cite{Li:2024kfm}.  This formula of Krylov complexity is valid for the thermal state and could be a criterion for assessing the accuracy of the wave function in the weak dissipative approximation. 

In a quantum system, Krylov entropy is the Shannon entropy of the probability distribution over the chosen Krylov basis. Following Ref.~\cite{Barbon:2019wsy}, we define it by
\begin{equation}
        S_K=-\sum^{\infty}_{n=0}|\psi_{n}|^{2}\ln|\psi_{n}|^{2},
        \label{k entropy}
    \end{equation}
where $\phi_n$ is operator wave function in Eq. \eqref{operator wave function of close syetem}. Then, the expression of Krylov entropy in the ideal closed system can be obtained by calculation as follows
\begin{equation}
    \begin{split}
    S_K&=-\sum^{\infty}_{n=0}|\psi_{n}|^{2}\ln|\psi_{n}|^{2}\\
    &=(K+1)\ln(K+1)-K\ln K.
    \end{split}
    \label{Krylov entropy of close system}
\end{equation}
Subsequently, we could investigate the Krylov entropy in three cases: the canonical single-field inflation, the non-trivial sound speed and the modified dispersion relation.

\subsection{Physical and cosmological implications}
\label{Physical and cosmological implications}

The quantities studied in this work are information-theoretic diagnostics. They are not direct cosmological observables. The effective temperature and squeezing angle describe the occupation and phase structure of the primordial modes. These quantities are determined by the same mode functions that enter cosmological correlation functions. However, this relation does not establish a direct observational signal. The non-trivial sound speed changes the mode frequency and the sound-horizon crossing condition. A periodic sound speed can also generate resonance in the mode evolution. The oscillations found in the thermal and Krylov quantities reflect this modulation. A quantitative connection with primordial black holes requires a separate calculation of the scalar power spectrum. The modified dispersion relation changes the mode evolution at high physical wavenumber. It can produce transient oscillations in the calculated Krylov quantities. The form of these oscillations depends on the selected dispersion function and its parameters.  
The present results therefore describe Krylov space response of the three models. They do not provide direct predictions for cosmological observations.

\subsection{Krylov complexity and Krylov entropy in canonical single-field inflation}
\label{sec:6.1}

In Section \ref{sec:canonical single-field inflation}, we already obtained the evolution of the effective temperature $T_k(\eta)$ in Eq. \eqref{eq:5.3} as follows
\begin{equation}\label{eq:evolution of Tk}
\frac{dT_k}{d\eta}=\frac{4k_{\rm B}T_k^{2}}{\omega}\frac{1}{\eta}\cos (2\phi_k)\sinh{\big(\frac{\omega}{2k_{\rm B}T_k}\big)}  
\end{equation}
Both Krylov complexity and Krylov entropy are predominantly governed by effective temperature. Therefore we substitute the temperature evolution data obtained under canonical single-field inflation \eqref{eq:evolution of Tk} into Eqs. \eqref{Krylov complexity of close system} and \eqref{Krylov entropy of close system} to acquire the evolutionary plots presented below. 
\begin{figure}[h!]
    \centering  
    \includegraphics[width=0.45\linewidth]{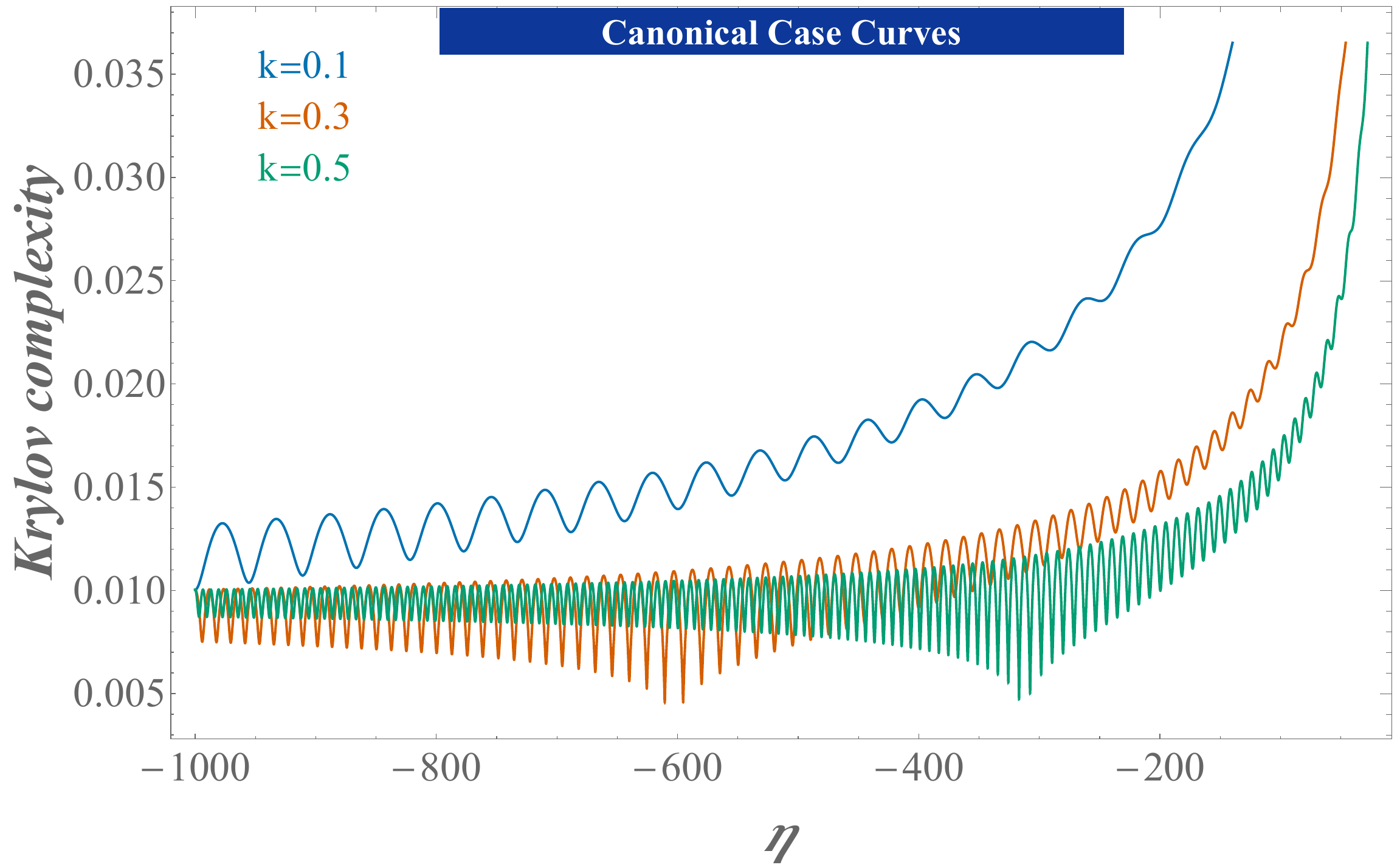}
    \includegraphics[width=0.45\linewidth]{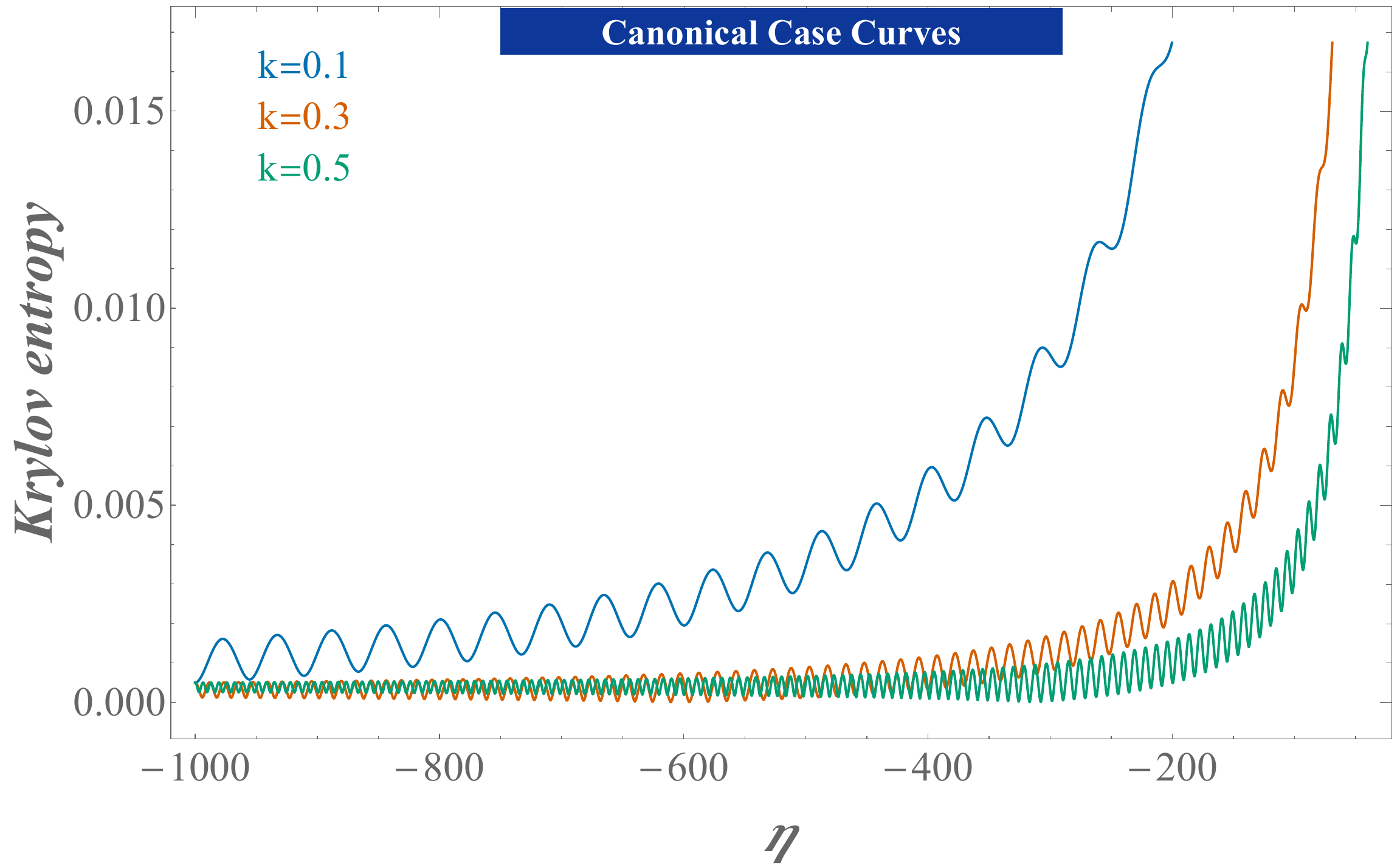}
    \caption{Krylov complexity and Krylov entropy as functions of $\eta$ for canonical single-field inflation. We use $k=0.1,0.3,0.5$, $\omega=1$, and $k_{\rm B}=1$.}
    \label{fig:Krylov complexity and Krylov entropy of canonical single-field inflation}
\end{figure}

In Fig. \ref{fig:Krylov complexity and Krylov entropy of canonical single-field inflation}, The left panel shows the evolution of Krylov complexity with conformal time $\eta$ ranging from -$1000$ to $0$ under canonical single-field inflation for different values of comoving-wavenumber $k$. In the early inflationary stage when $\eta$ changes from -$1000$ to -$400$ all curves produce small high frequency oscillations around separate baseline values. At fixed conformal time, the $k = 0.1$ mode has the largest Krylov complexity in the plotted interval. Because different wavenumbers cross the horizon at different times, a physically cleaner comparison is obtained by plotting the curves against $x = -k\eta$ or by marking the individual horizon-crossing times. After $\eta$ exceeds -$400$ and gradually approaches $0$ the steady oscillatory behaviour terminates and Krylov complexity enters a phase of nonlinear accelerated growth. The growth rate rises sharply near the end of inflation as $\eta$ tends to $0$ and numerical discrepancies between different $k$ become increasingly notable. Larger comoving-wavenumber restricts the accumulation of Krylov complexity by lowering early baseline values and postponing the onset of rapid Krylov complexity growth. The right panel describes Krylov entropy evolution over $\eta$ from -$1000$ to $0$ using the same set of momentum parameters and shares nearly identical overall trends with Krylov complexity. Within the interval from -$1000$ to -$400$ entropy values for $k=0.3$ and $k=0.5$ remain close to zero with trivial fluctuations while $k=0.1$ maintains a much higher Krylov entropy baseline. All Krylov entropy curves start steep growth once $\eta$ passes -$400$ and the upward trend intensifies when $\eta$ approaches $0$ with smaller momentum generating larger final Krylov entropy. This behavior indicates a rapid increase in the spread of the thermal state over the chosen Krylov basis.

\subsection{Krylov complexity and Krylov entropy of non-trivial sound speed}
\label{sec:Krylov complexity and Krylov entropy of non-trivial sound speed}

We present the dynamical evolution equation for temperature within the non-trivial sound speed \eqref{eq:5.6} 
\begin{equation} \label{Tk of non-trivial sound speed}
 \frac{dT_k}{d\eta}=\frac{4k_{\rm B}T_k^{2}}{\omega}\sinh{\frac{\omega}{2k_{\rm B}T_k}}\bigg[\frac{1}{\eta}\cos{(2\phi_k)}+\frac{k}{2}(c_s^2-1)\sin{(2\phi_k)}\bigg], 
\end{equation}
being armed with these two Eqs. \eqref{Krylov complexity of close system} and \eqref{Krylov entropy of close system}, we could utilize their numeric to simulate the evolution of Krylov complexity and Krylov entropy in the non-trivial sound speed model. 
 \begin{figure}
        \centering
        \includegraphics[width=0.45\linewidth]{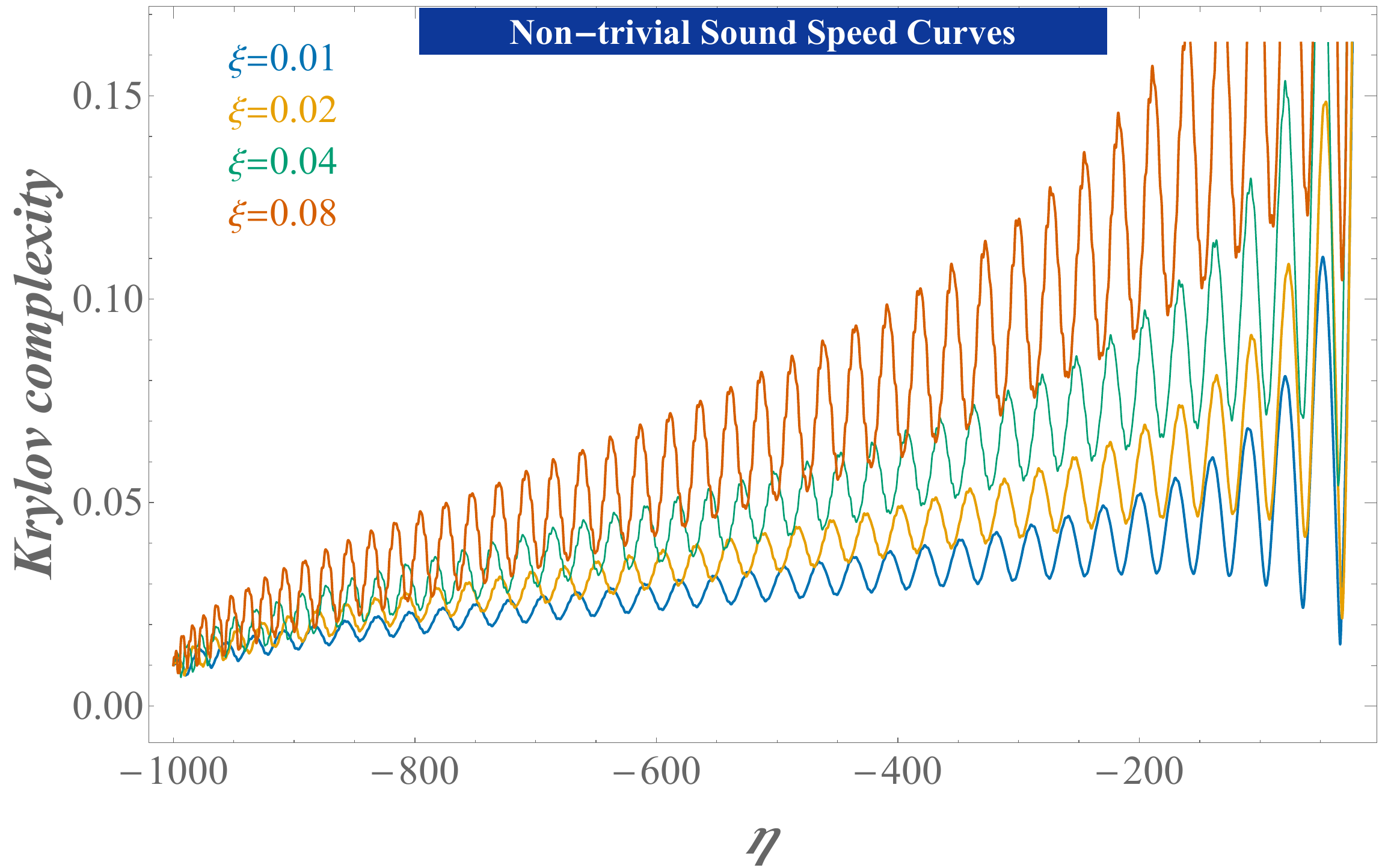} 
        \includegraphics[width=0.45\linewidth]{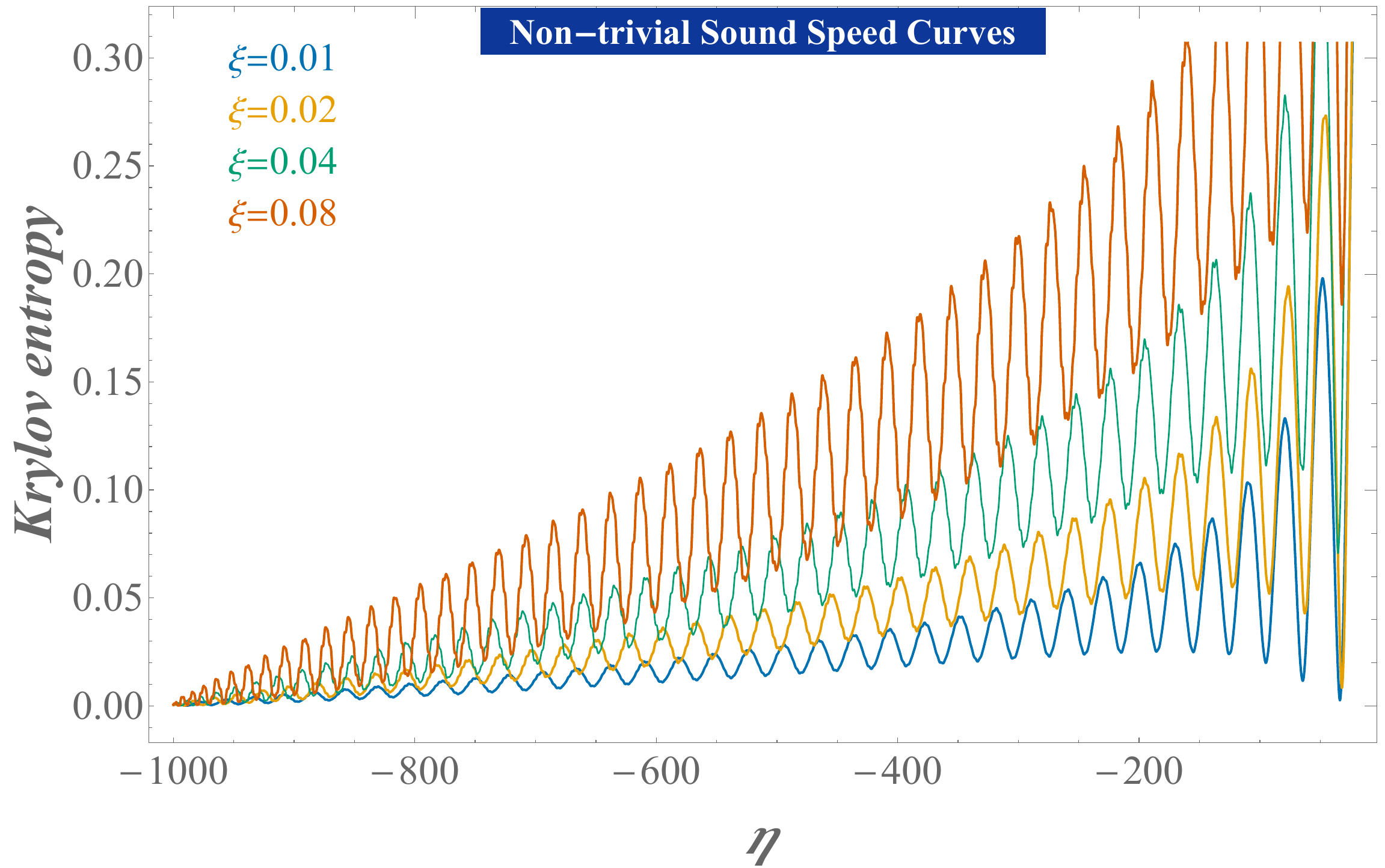}
        \caption{Krylov complexity and Krylov entropy as functions of $\eta$ for the non-trivial sound speed model. We use $k=0.1$, $\omega=1$, $k_{\rm B}=1$, and $\xi=0.01,0.02,0.04,0.08$.}
        \label{fig:Krylov complexity and Krylov entropy of non-trivial sound speed}
    \end{figure}

Fig. \ref{fig:Krylov complexity and Krylov entropy of non-trivial sound speed} displays the Krylov complexity and Krylov entropy of non-trivial sound speed varying with $\xi$ and $T_k$. The left set of curves illustrates how Krylov complexity evolves with conformal time $\eta$ running from $-1000$ to $0$ under the non-trivial sound speed framework. Fixed parameters $k=0.1$ and $\omega=1$ are used throughout calculations while four non-trivial sound speed correction values $\xi=0.01,0.02,0.04,0.08$ are selected for comparative simulation. Krylov complexity begins near zero with mild oscillations in the early evolutionary stage. It keeps increasing alongside growing oscillation amplitude as conformal time approaches the late inflationary epoch. Larger non-trivial sound speed correction parameters yield higher mean Krylov complexity and larger oscillation peaks at the same time slice. Stronger sound speed corrections improve dynamical coupling between field modes accelerate operator diffusion inside Hilbert space and produce higher accumulated Krylov complexity values. The numerical gaps between curves of different $\xi$ expand near the end of inflation and the boosting effect of non-trivial sound speed on complexity growth becomes much more prominent. 

The second set of curves is computed using the same group of non-trivial sound speed parameters to obtain Krylov entropy evolution against conformal time. Krylov entropy stays close to zero with faint oscillations at early evolution. Its average value and oscillation intensity rise steadily as cosmic evolution proceeds. Larger $\xi$ leads to higher instantaneous entropy and broader oscillation amplitude. From physical perspective sound speed correction introduces greater disorder into quantum state evolution and spreads probability distribution over the Krylov basis. Krylov entropy differences among distinct parameter configurations become more obvious toward late inflation. The non-trivial sound speed correction can adjust the growth rate and final magnitude of both physical quantities simultaneously.

\subsection{Krylov complexity and Krylov entropy of modified dispersion relation}
\label{sec:Krylov complexity and Krylov entropy of modified dispersion relation}

Following the similar procedure, we could rewrite the evolution of $T_k$ 
\begin{equation}\label{eq:Tk of modified dispersion relation}
  \frac{dT_k}{d\eta}=\frac{4k_{\rm B}T_k^{2}}{\omega}\sinh{\frac{\omega}{2k_{\rm B}T_k}}\bigg[\frac{1}{\eta}\cos{(2\phi_k)}+\frac{k}{2}(f^2-1)\sin{(2\phi_k)}\bigg].      
\end{equation}
By substituting Eq. \ref{eq:Tk of modified dispersion relation} into Eqs. \ref{Krylov complexity of close system} and \ref{Krylov entropy of close system}, we further analyze the evolution of Krylov complexity and Krylov entropy within the modified dispersion relation model. 
    \begin{figure}[h!]
        \centering
        \includegraphics[width=0.45\linewidth]{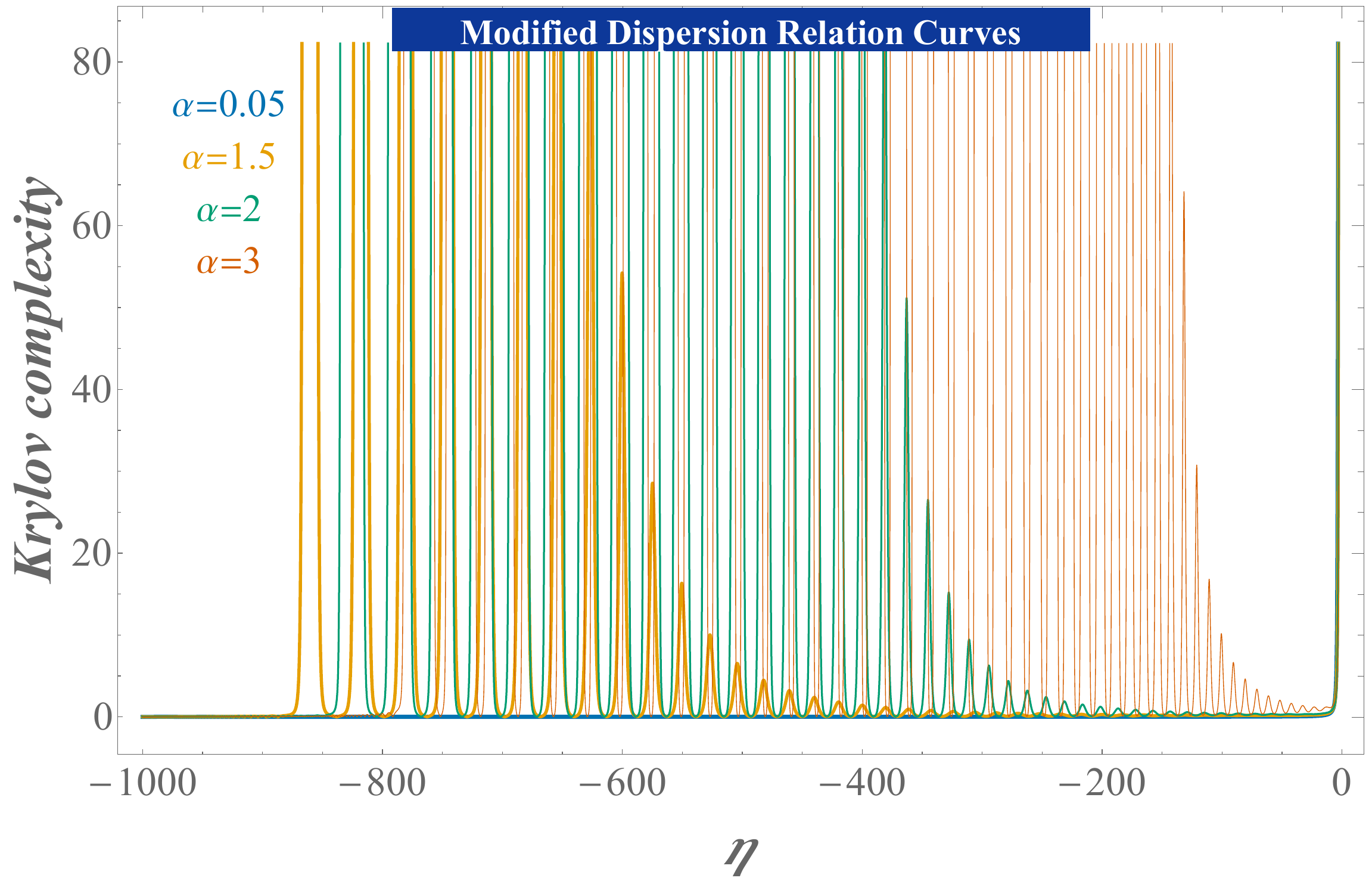} 
        \includegraphics[width=0.45\linewidth]{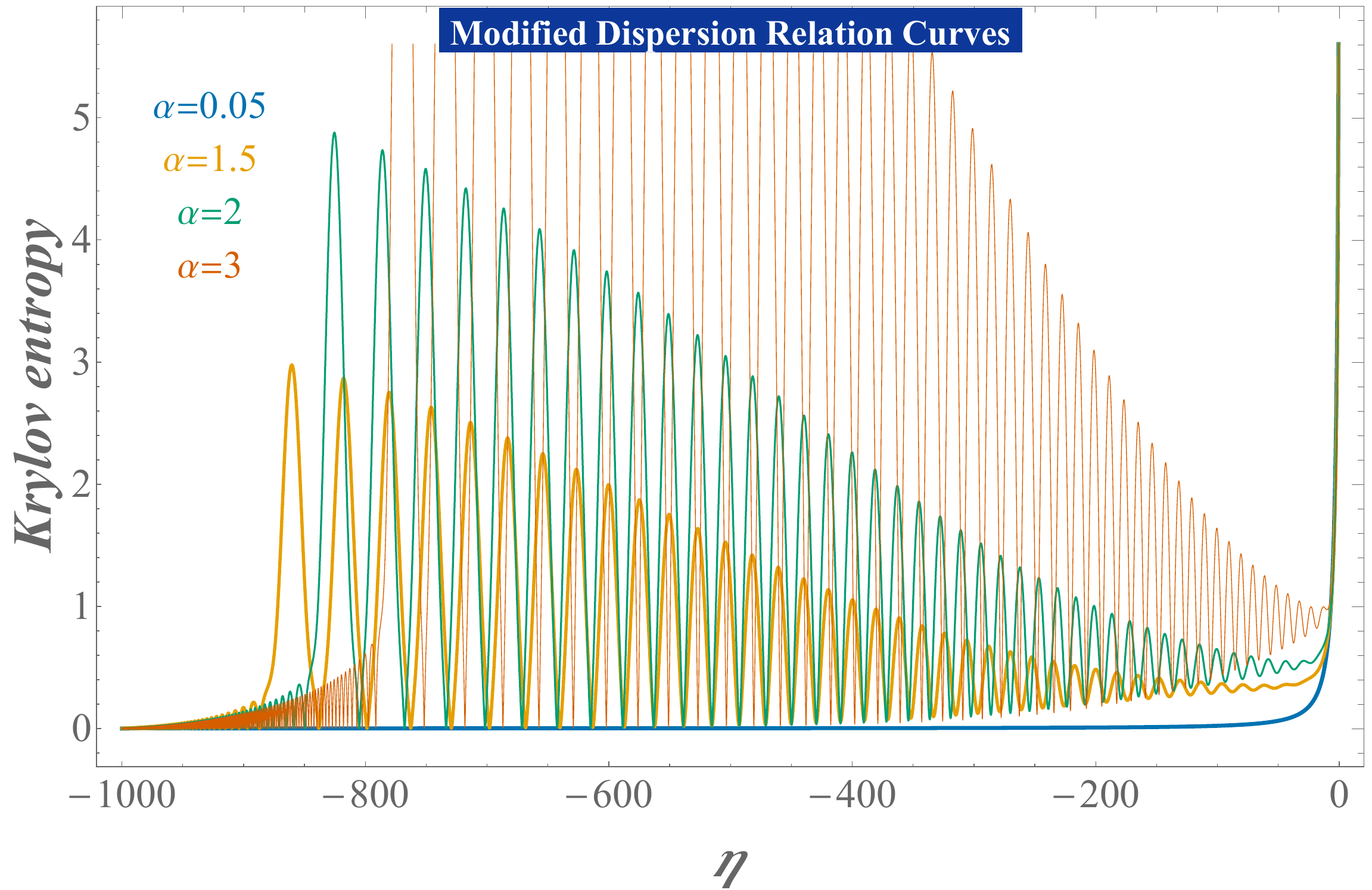}
        \caption{Krylov complexity and Krylov entropy as functions of $\eta$ for the modified dispersion relation. We use $k=0.1$, $k_{\rm B}=1$, $\omega=1$, $\alpha=0.05,1.5,3.0$, $M=1$, and $k_{\rm ph}/M=1.5$.}
        \label{fig:Krylov complexity and Krylov entropy of modified dispersion relation}
    \end{figure}

Fig. \ref{fig:Krylov complexity and Krylov entropy of modified dispersion relation} illustrates the evolution of Krylov complexity and Krylov entropy with respect to varying values of $\alpha$ and $T_k$ for the modified dispersion relation. These left curves present the evolution of Krylov complexity with conformal time $\eta$ ranging from -$1000$ to $0$ under the modified dispersion relation framework. Three dispersion correction parameters $\alpha=0.05,1.5,3$ are adopted for comparative simulation. Krylov complexity stays near zero with negligible fluctuations during the early evolutionary stage. Intense periodic oscillations emerge after conformal time crosses certain threshold value. Larger modified dispersion relation leads to earlier oscillation onset and higher peak value within each oscillation cycle. Oscillation peaks decrease gradually in the mid evolutionary stage while cases with stronger modified dispersion relation retain larger oscillation amplitude. All Krylov complexity curves undergo rapid final growth as $\eta$ approaches zero near the end of inflation. Enhanced modified dispersion relation strengthens dynamical interference between field modes and accelerates operator diffusion across Hilbert space to produce stronger Krylov complexity oscillations. Persistent numerical gaps among different parameter sets show that the modified dispersion relation continuously adjusts oscillation onset peak magnitude and late growth behaviour of Krylov complexity. Corresponding Krylov entropy results are calculated with identical dispersion correction parameters and share highly consistent evolutionary patterns with Krylov complexity. Krylov entropy remains close to zero without obvious fluctuations at early times. Pronounced oscillations in Krylov entropy start after the same critical conformal time and stronger modified dispersion relation yields earlier oscillation emergence along with larger peak values. Peak Krylov entropy declines gradually in mid evolution and the separation between curves for different parameters becomes increasingly visible. The modified dispersion relation raises the disorder of quantum evolution and broadens probability distribution defined on Krylov basis to generate higher entropy. All Krylov entropy curves rise again near the end of inflation. The modified dispersion relation modulates the oscillatory properties and final values of both physical quantities simultaneously.

In Sec.~\ref{krylov complexity and Krylov entropy in closed system}, we calculate Krylov complexity and Krylov entropy in the three models. In canonical single-field inflation, the comoving-wavenumber changes their late-time growth. The non-trivial sound speed changes their mean values and oscillation amplitudes. The modified dispersion relation produces a different stage-dependent oscillation pattern. These results describe how the selected mode functions change spreading in the chosen Krylov basis.

\section{Effective open system description of Krylov dynamics}
\label{krylov complexity and Krylov entropy in open system}

The two-mode thermal state is a purification and evolves unitarily in the enlarged Hilbert space. It does not by itself describe irreversible damping. We therefore introduce a separate phenomenological generator for the selected Krylov sector. This construction follows Refs.~\cite{Bhattacharya:2022gbz,Bhattacharya:2023zqt,Li:2024kfm}. It is an effective parametrization rather than a microscopic Lindblad model.

\subsection{Effective Krylov coefficients}
\label{sec:Lanczos coefficient and dissipation coefficient}

The generator acts on the paired Krylov basis as
\begin{align}
&\mathcal L_{\rm eff}|\mathcal O_n)=-i\gamma_n|\mathcal O_n)+b_{n+1}|\mathcal O_{n+1})+b_n|\mathcal O_{n-1}),
\label{eq:9.2}\\
b_n^2&=\mathcal A_k n(n-1+\nu),\quad
\gamma_n=\mathcal B_k(2n+\nu),\quad \nu=1.
\label{eq:9.10}
\end{align}
Here $\mathcal A_k$ controls coherent spreading. The coefficient $\mathcal B_k$ controls the effective attenuation in the adopted ansatz. The three models are summarized by
\begin{equation}
(\mathcal A_k,\mathcal B_k)=
\begin{cases}
\left[(a'/a)^2,\ k\right], & \text{canonical single-field inflation},\\[2pt]
\left[\left(\frac{k}{2}(c_s^2-1)\right)^2+(a'/a)^2,\ \frac{k}{2}(c_s^2+1)\right], & \text{non-trivial sound speed},\\[2pt]
\left[\left(\frac{k}{2}(f^2-1)\right)^2+(a'/a)^2,\ \frac{k}{2}(f^2+1)\right], & \text{modified dispersion relation}.
\end{cases}
\label{mu for the three inflationary models}
\end{equation}
For $\nu=1$, direct substitution gives the coefficients that enter the effective generator,
\begin{equation}
\begin{aligned}
&b_n^{\rm can}=n\left|\frac{a'}{a}\right|,
&\gamma_n^{\rm can}&=k(2n+1),\\
&b_n^{c_s}=n\sqrt{\left[\frac{k}{2}(c_s^2-1)\right]^2+\left(\frac{a'}{a}\right)^2},
&\gamma_n^{c_s}&=\frac{k}{2}(c_s^2+1)(2n+1),\\
&b_n^{\rm MDR}=n\sqrt{\left[\frac{k}{2}(f^2-1)\right]^2+\left(\frac{a'}{a}\right)^2},
&\gamma_n^{\rm MDR}&=\frac{k}{2}(f^2+1)(2n+1).
\end{aligned}
\label{eq:explicit open coefficients}
\end{equation}
The sound-speed model can be written entirely in terms of $\xi$. Equation~\eqref{Non-trivial sound speed with k} gives
\begin{equation}
\mathcal A_k^{c_s}=k^2\xi^2\left[1-\cos(k_\ast\eta)\right]^2+\left(\frac{a'}{a}\right)^2,
\qquad
\mathcal B_k^{c_s}=k\left\{1-\xi\left[1-\cos(k_\ast\eta)\right]\right\}.
\label{eq:explicit sound coefficients}
\end{equation}
The modified dispersion relation gives
\begin{equation}
(\mathcal A_k^{\rm MDR},\mathcal B_k^{\rm MDR})=
\begin{cases}
\left[\dfrac{k^2}{4}\left\{\left(\dfrac{k_{\rm ph}}{M}\right)^{2\alpha}-1\right\}^2+
\left(\dfrac{a'}{a}\right)^2,
\dfrac{k}{2}\left\{\left(\dfrac{k_{\rm ph}}{M}\right)^{2\alpha}+1\right\}\right],
& k_{\rm ph}>M,\\[6pt]
\left[\left(\dfrac{a'}{a}\right)^2,k\right],&k_{\rm ph}\leq M.
\end{cases}
\label{eq:explicit MDR coefficients}
\end{equation}
Below the scale $M$, the modified dispersion relation therefore reduces to the canonical coefficients. In the quasi-de Sitter approximation, $a'/a\simeq-1/\eta$. The common background contribution to $\mathcal A_k$ is then $1/\eta^2$. The linear dependence of $b_n$ on $n$ follows from the two-mode algebra. It characterizes spreading in the selected Krylov basis. It does not by itself prove maximal quantum chaos \cite{Parker:2018yvk}.

\begin{figure}[htbp]
\centering
\begin{minipage}{0.32\textwidth}
\centering
\includegraphics[width=\linewidth]{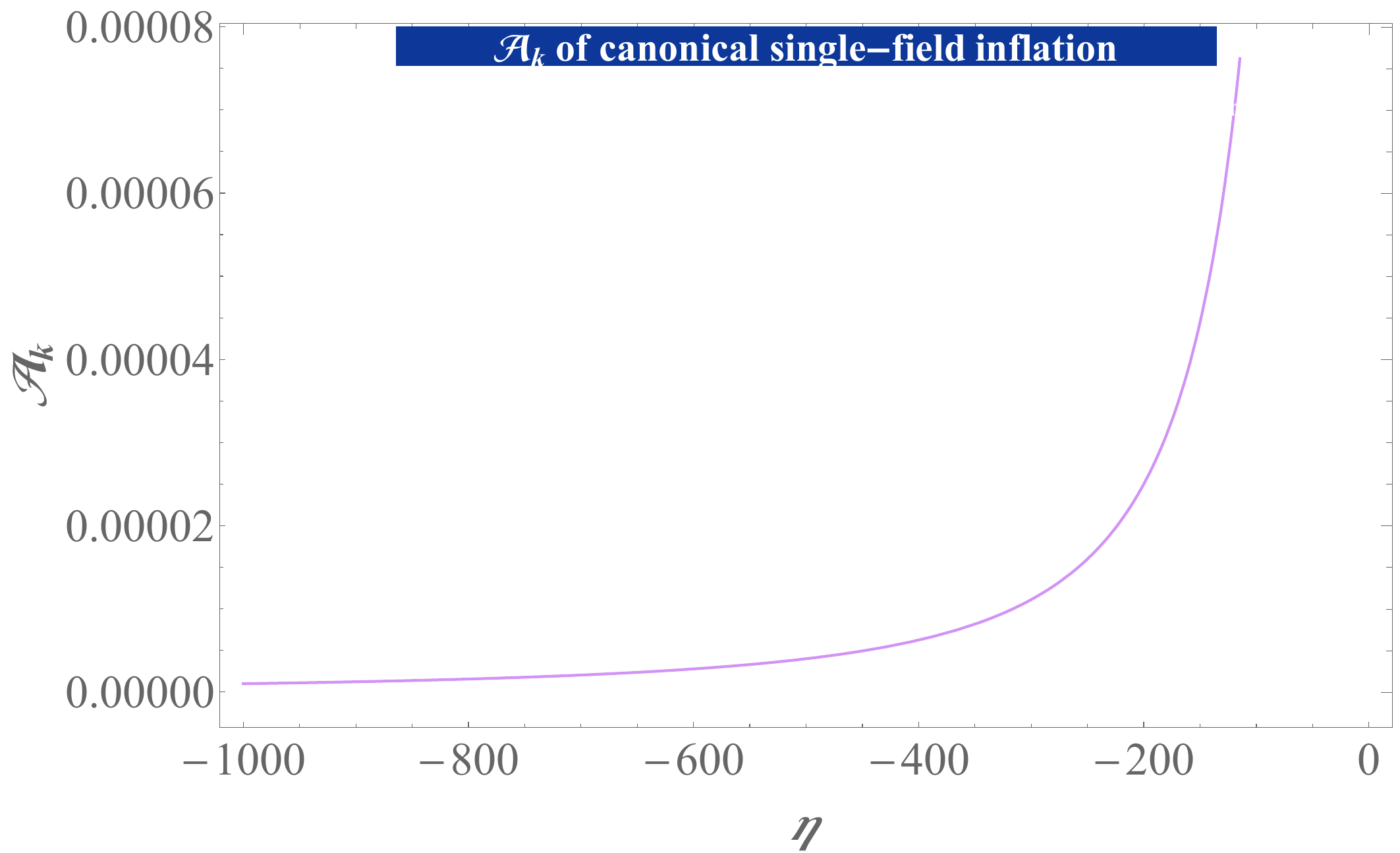}
\end{minipage}\hfill
\begin{minipage}{0.32\textwidth}
\centering
\includegraphics[width=\linewidth]{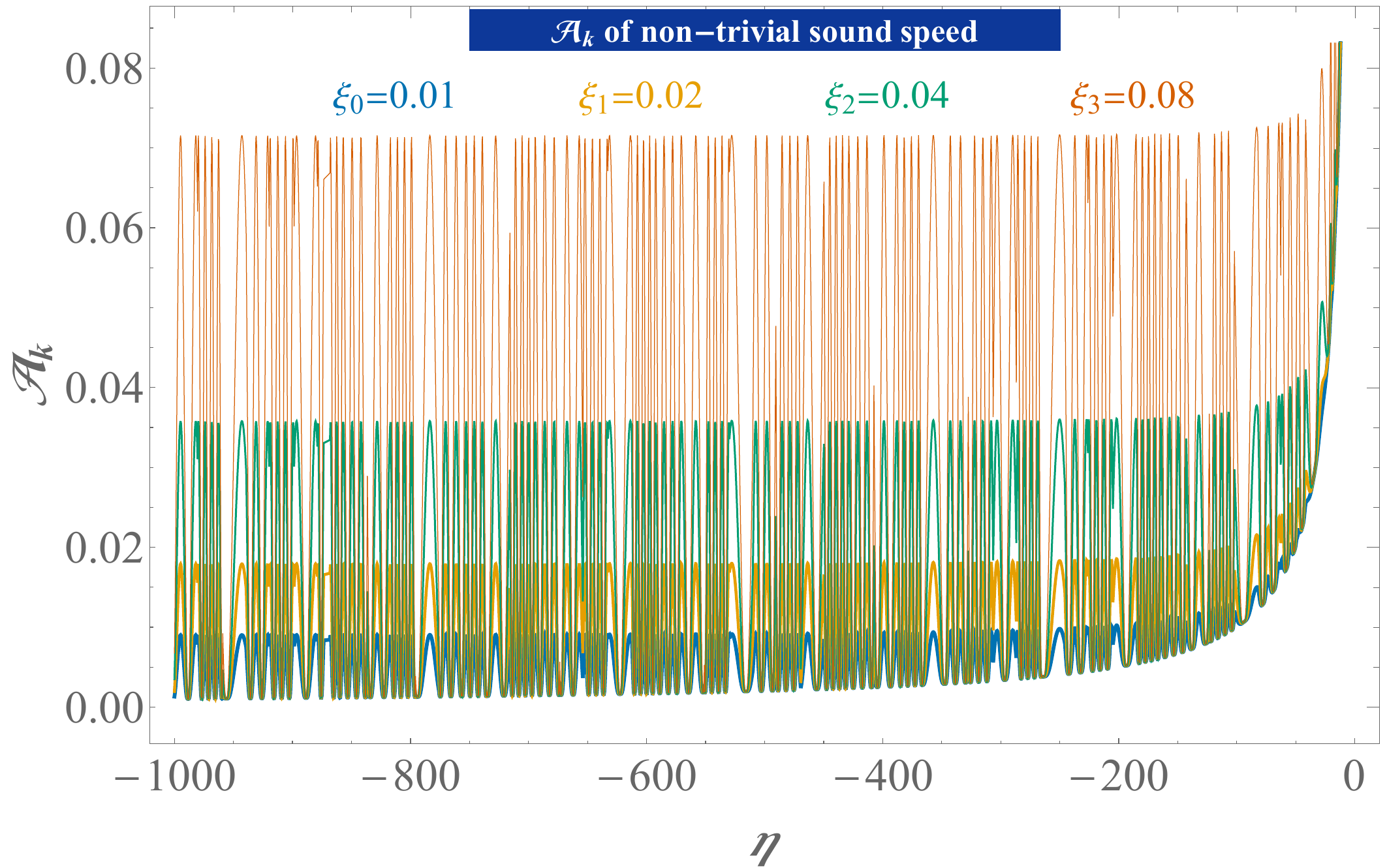}
\end{minipage}\hfill
\begin{minipage}{0.32\textwidth}
\centering
\includegraphics[width=\linewidth]{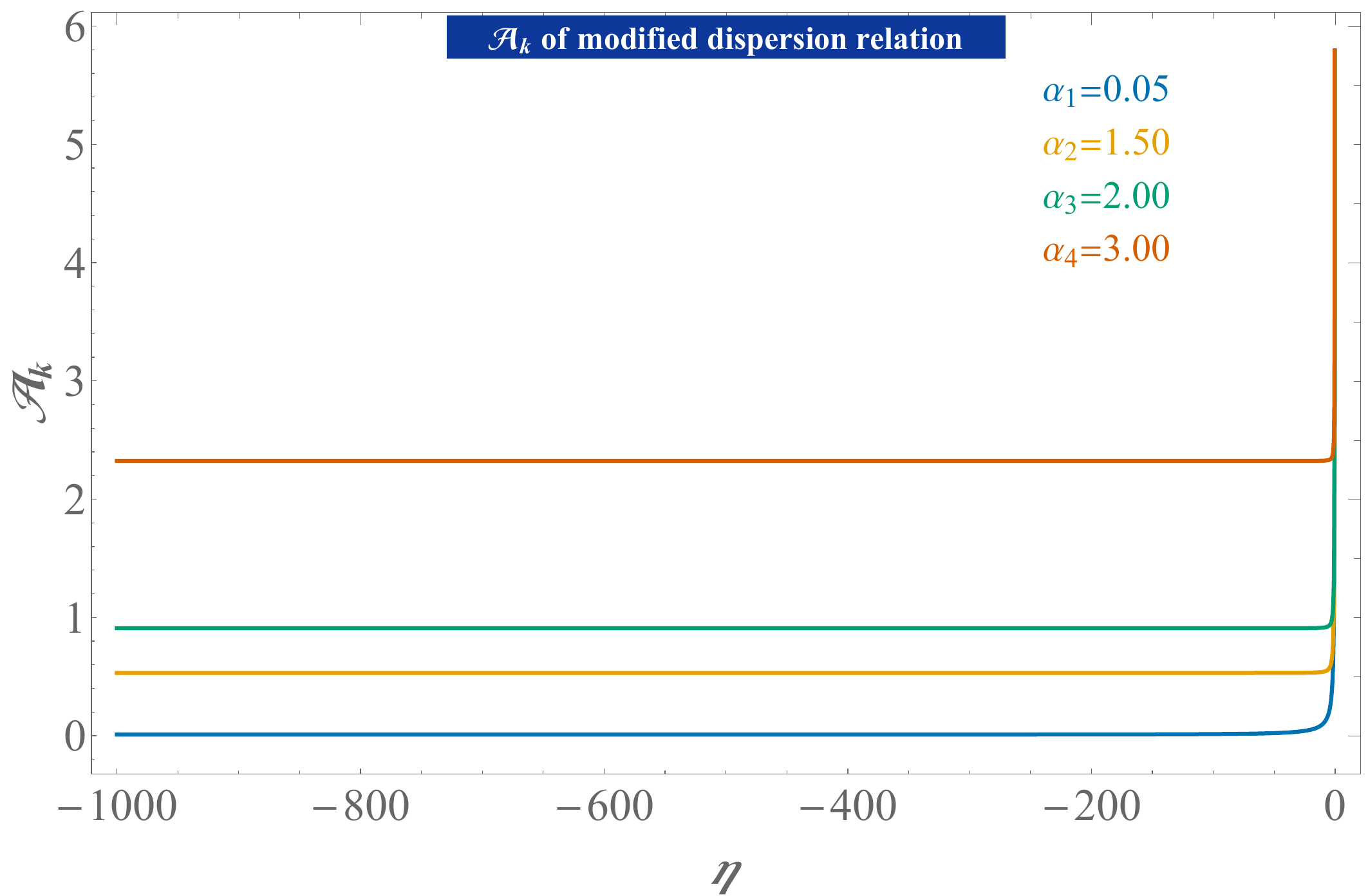}
\end{minipage}
\caption{The coherent-spreading coefficient \(\mathcal A_k\) in canonical single-field inflation, the non-trivial sound speed model, and the modified dispersion relation model. The canonical panel uses \(\mathcal A_k=(a'/a)^2\). The sound-speed panel includes the periodic contribution \(k^2\xi^2[1-\cos(k_\ast\eta)]^2\). The modified-dispersion-relation panel includes the ultraviolet contribution determined by \(f(k_{\rm ph})\). We set \(k=0.1\), \(\xi=0.01,0.02,0.04,0.08\), and \(\alpha=0.05,1.50,2.00,3.00\). The values of \(k_\ast\), \(M\), and \(k_{\rm ph}/M\) are specified in the numerical calculation. }
\label{fig:bn of three models}
\end{figure}

Fig.~\ref{fig:bn of three models} compares the coherent-spreading coefficient \(\mathcal A_k\) in the three models. In canonical single-field inflation, \(\mathcal A_k=(a'/a)^2\simeq1/\eta^2\). It therefore remains small at early times and increases as \(\eta\) approaches zero. In the non-trivial sound speed model, the additional term \(k^2\xi^2[1-\cos(k_\ast\eta)]^2\) produces a periodic modulation. Its amplitude increases with \(\xi\). In the modified dispersion relation model, the ultraviolet term produces an \(\alpha\)-dependent offset. The common background term \(1/\eta^2\) becomes important near the end of the plotted interval. These differences show how the three wavenumber dependences modify coherent spreading in the selected Krylov basis. 

\section{Normalized thermal Krylov observables in the effective open system}
\label{krylov complexity in open system}
\label{k entropy in open system}

We now derive the normalized thermal Krylov observables for a general mode temperature $T_k$. The temperature enters through the thermal occupation factor $q_k$. The model dependence enters through the effective coefficients $\mathcal A_k$ and $\mathcal B_k$. This separation allows the thermal contribution and the wavenumber-dependent correction to be identified directly.
\begin{equation}
\psi_n=\frac{\sqrt{1-q_k}}{1+\mathcal B_k\sqrt{q_k}}\,
\mathcal A_k^{n/2}
\left(\frac{\sqrt{q_k}}{1+\mathcal B_k\sqrt{q_k}}\right)^n,
\qquad q_k=e^{-\omega/(k_{\rm B}T_k)}.
\label{wave function for open}
\end{equation}
The amplitudes are not normalized in general. We therefore define the geometric ratio, the norm, and the normalized probability in one step,
\begin{align}
R_k&=\frac{\mathcal A_k q_k}{(1+\mathcal B_k\sqrt{q_k})^2},\qquad 0\leq R_k<1,
\label{eq:9.14}\\
\mathcal N_k&=\sum_{n=0}^{\infty}|\psi_n|^2
=\frac{1-q_k}{(1+\mathcal B_k\sqrt{q_k})^2(1-R_k)},\qquad
p_n=\frac{|\psi_n|^2}{\mathcal N_k}=(1-R_k)R_k^n.
\label{eq:9.7}
\end{align}
The condition $R_k<1$ ensures normalizability. The normalized Krylov complexity and Krylov entropy follow from the same geometric distribution,
\begin{align}
K&=\sum_{n=0}^{\infty}np_n=\frac{R_k}{1-R_k},
\label{eq:9.15}\\
S_K&=-\sum_{n=0}^{\infty}p_n\ln p_n
=-\ln(1-R_k)-\frac{R_k}{1-R_k}\ln R_k.
\label{eq:11.2}
\end{align}
In the ideal thermal limit $\mathcal A_k\to1$ and $\mathcal B_k\to0$, Eq.~\eqref{eq:9.15} reduces to
\begin{equation}
K\longrightarrow\frac{q_k}{1-q_k}
=\frac{1}{e^{\omega/(k_{\rm B}T_k)}-1}.
\label{leading order of krylov complexity}
\end{equation}

We now insert the coefficients of the three models. Define
\begin{equation}
D_X=\left(1+\mathcal B_k^X\sqrt{q_k}\right)^2-\mathcal A_k^Xq_k,
\qquad X\in\{{\rm can},c_s,{\rm MDR}\}.
\label{eq:open denominators}
\end{equation}
It follows that $R_k^X=\mathcal A_k^Xq_k/(1+\mathcal B_k^X\sqrt{q_k})^2$ and $K_X=\mathcal A_k^Xq_k/D_X$. The three results are
\begin{align}
&K_{\rm can}=
\frac{(a'/a)^2q_k}{(1+k\sqrt{q_k})^2-(a'/a)^2q_k},
\label{eq:open K canonical}\\
&K_{c_s}=
\frac{\left\{k^2\xi^2[1-\cos(k_\ast\eta)]^2+(a'/a)^2\right\}q_k}
{\left\{1+k[1-\xi(1-\cos(k_\ast\eta))]\sqrt{q_k}\right\}^2
-\left\{k^2\xi^2[1-\cos(k_\ast\eta)]^2+(a'/a)^2\right\}q_k},
\label{eq:open K sound}\\
&K_{\rm MDR}=
\begin{cases}
\dfrac{\left[\dfrac{k^2}{4}(F_\alpha-1)^2+(a'/a)^2\right]q_k}
{\left[1+\dfrac{k}{2}(F_\alpha+1)\sqrt{q_k}\right]^2
-\left[\dfrac{k^2}{4}(F_\alpha-1)^2+(a'/a)^2\right]q_k},
&k_{\rm ph}>M,\\[10pt]
\dfrac{(a'/a)^2q_k}{(1+k\sqrt{q_k})^2-(a'/a)^2q_k},
&k_{\rm ph}\leq M,
\end{cases}
\qquad F_\alpha=\left(\frac{k_{\rm ph}}{M}\right)^{2\alpha}.
\label{eq:open K MDR}
\end{align}
The second branch of Eq.~\eqref{eq:open K MDR} is identical to Eq.~\eqref{eq:open K canonical}. This is the direct consequence of $f=1$ for $k_{\rm ph}\leq M$. In the quasi-de Sitter approximation, Eq.~\eqref{eq:open K canonical} can also be written as $K_{\rm can}=q_k/[\eta^2(1+k\sqrt{q_k})^2-q_k]$. The corresponding entropy does not require a new summation. For each model, it is
\begin{equation}
S_K^X=(K_X+1)\ln(K_X+1)-K_X\ln K_X,
\qquad X\in\{{\rm can},c_s,{\rm MDR}\}.
\label{eq:open entropy three models}
\end{equation}
Eqs.~\eqref{eq:open K canonical}--\eqref{eq:open entropy three models} are the explicit open-system results. Both observables are determined by the same geometric ratio $R_k$. Their dependence on the thermal factor and the effective coefficients is
\begin{align}
\frac{\partial R_k}{\partial\mathcal A_k}
&=\frac{q_k}{\left(1+\mathcal B_k\sqrt{q_k}\right)^2}>0,
&
\frac{\partial R_k}{\partial\mathcal B_k}
&=-\frac{2\mathcal A_kq_k^{3/2}}
{\left(1+\mathcal B_k\sqrt{q_k}\right)^3}<0,
\nonumber\\
\frac{\partial R_k}{\partial q_k}
&=\frac{\mathcal A_k}
{\left(1+\mathcal B_k\sqrt{q_k}\right)^3}>0,
&
\frac{\partial q_k}{\partial T_k}
&=\frac{\omega q_k}{k_{\rm B}T_k^2}>0,
\nonumber\\
\frac{\partial K}{\partial R_k}
&=\frac{1}{(1-R_k)^2}>0,
&
\frac{\partial S_K}{\partial R_k}
&=-\frac{\ln R_k}{(1-R_k)^2}>0.
\label{eq:9.8}
\end{align}
These relations hold in the normalizable region $0<R_k<1$. The coefficient $\mathcal A_k$ enhances spreading in the selected Krylov basis. A positive $\mathcal B_k$ reduces the geometric ratio. At fixed conformal time and fixed model parameters, an increase in $T_k$ increases $q_k$. It therefore increases both the normalized thermal Krylov complexity and the Krylov entropy.

The thermal limits make this behavior more explicit. At low temperature, $q_k\rightarrow0$. The leading terms are
\begin{equation}
R_k=\mathcal A_kq_k+\mathcal O(q_k^{3/2}),
\qquad
K=\mathcal A_kq_k+\mathcal O(q_k^{3/2}),
\qquad
S_K=R_k(1-\ln R_k)+\mathcal O(R_k^2).
\label{eq:open low temperature}
\end{equation}
The probability distribution is then concentrated near the first Krylov basis state. The thermal Krylov complexity and the Krylov entropy are suppressed by the Boltzmann factor.

At high temperature, $k_{\rm B}T_k\gg\omega$, and
\begin{equation}
q_k=1-\frac{\omega}{k_{\rm B}T_k}+\mathcal O(T_k^{-2}).
\end{equation}
Substitution into Eq.~\eqref{eq:9.14} gives
\begin{equation}
R_k=\frac{\mathcal A_k}{(1+\mathcal B_k)^2}
\left[1-\frac{\omega}
{k_{\rm B}T_k(1+\mathcal B_k)}+\mathcal O(T_k^{-2})\right].
\label{eq:open high temperature}
\end{equation}
The temperature dependence is therefore weak in this regime. In the formal infinite-temperature limit,
\begin{equation}
R_k^{(\infty)}=\frac{\mathcal A_k}{(1+\mathcal B_k)^2},
\qquad
K^{(\infty)}=\frac{\mathcal A_k}
{(1+\mathcal B_k)^2-\mathcal A_k},
\label{eq:open infinite temperature}
\end{equation}
provided that $\mathcal A_k<(1+\mathcal B_k)^2$. These expressions describe the thermal saturation of the normalized distribution. They also show why the temperature dependence becomes small when $T_k$ is much larger than $\omega/k_{\rm B}$.

For canonical single-field inflation, the ratio is
\begin{equation}
R_k^{\rm can}=\frac{(a'/a)^2q_k}{(1+k\sqrt{q_k})^2}.
\end{equation}
In the quasi-de Sitter approximation, $(a'/a)^2\simeq1/\eta^2$. The background contribution grows as $\eta$ approaches zero. Hence, $R_k^{\rm can}$, $K_{\rm can}$, and $S_K^{\rm can}$ increase while the normalizability condition remains satisfied.

For the non-trivial sound speed model, define $C_k(\eta)=1-\cos(k_\ast\eta)$. The corresponding ratio is
\begin{equation}
R_k^{c_s}=\frac{\left[(a'/a)^2+k^2\xi^2C_k^2\right]q_k}
{\left[1+k(1-\xi C_k)\sqrt{q_k}\right]^2}.
\end{equation}
For positive $\xi$ in the range adopted here, the correction increases $\mathcal A_k^{c_s}$ and reduces $\mathcal B_k^{c_s}$. Both changes increase $R_k^{c_s}$ when $C_k\neq0$. The periodic factor $C_k$ produces an oscillatory modulation. The canonical result is recovered when $\xi=0$ or $C_k=0$.

For the modified dispersion relation, the canonical result is recovered exactly when $k_{\rm ph}\leq M$. When $k_{\rm ph}>M$, the ratio becomes
\begin{equation}
R_k^{\rm MDR}=\frac{\left[(a'/a)^2+\dfrac{k^2}{4}(F_\alpha-1)^2\right]q_k}
{\left[1+\dfrac{k}{2}(F_\alpha+1)\sqrt{q_k}\right]^2},
\qquad
F_\alpha=\left(\frac{k_{\rm ph}}{M}\right)^{2\alpha}.
\end{equation}
The modified dispersion relation increases both $\mathcal A_k^{\rm MDR}$ and $\mathcal B_k^{\rm MDR}$. The first change enhances $R_k^{\rm MDR}$. The second change suppresses it. The net change is determined by their competition. Its sign can be tested from
\begin{equation}
\frac{\partial\ln R_k^{\rm MDR}}{\partial F_\alpha}
=\frac{k^2(F_\alpha-1)}{2\mathcal A_k^{\rm MDR}}
-\frac{k\sqrt{q_k}}
{1+\mathcal B_k^{\rm MDR}\sqrt{q_k}}.
\label{eq:open MDR criterion}
\end{equation}
A positive value means that the coherent correction is dominant. A negative value means that the effective attenuation is dominant. This criterion avoids assigning a universal enhancement or suppression to the modified dispersion relation.

The three models therefore differ through $\mathcal A_k$ and $\mathcal B_k$. The thermal occupation is described by the common factor $q_k$. The normalized ratio $R_k$ combines these effects. It determines both the thermal Krylov complexity and the Krylov entropy. The analytical expressions provide the complete result within the adopted effective open-system parametrization. No additional open-system complexity figure is required.

\section{Summary}
\label{summary}

In this work, we studied thermal Krylov complexity and Krylov entropy for primordial scalar modes in three inflationary models. These models are canonical single-field inflation, a non-trivial sound speed model, and a modified dispersion relation model. They contain different comoving-wavenumber dependences. Canonical single-field inflation contains the standard \(k^2\) term. The non-trivial sound speed model replaces this term by \(c_s^2(\eta)k^2\). The modified dispersion relation introduces an additional dependence on \(k_{\rm ph}/M\).

We first constructed a purified two-mode thermal state. This state evolves unitarily in an enlarged Hilbert space. We used it to calculate the effective mode temperature and squeezing angle. We then studied thermal Krylov complexity and Krylov entropy in an ideal closed system. In canonical single-field inflation, the comoving wavenumber changes the effective mode temperature, the squeezing angle, and the late-time growth of the Krylov quantities. In the non-trivial sound speed model, the periodic sound-speed correction produces oscillatory modulation. Within the parameter range considered here, increasing \(\xi\) increases the mean values and oscillation amplitudes. The modified dispersion relation produces a different stage-dependent oscillatory structure.

We then introduced an independent effective open-system parametrization. The normalized Krylov distribution is controlled by the geometric ratio \(R_k\). The coefficient \(\mathcal A_k\) describes coherent Krylov-space spreading. The coefficient \(\mathcal B_k\) describes effective attenuation. At low temperature, the probability distribution is concentrated near the initial Krylov basis state. Both Krylov complexity and Krylov entropy are suppressed by the Boltzmann factor. At high temperature, their explicit temperature dependence becomes weak at fixed \(\mathcal A_k\) and \(\mathcal B_k\). A finite limiting distribution is obtained when the normalizability condition is satisfied.

The sound-speed correction increases \(R_k\) in the parameter range adopted here. The modified dispersion relation has a different effect. It increases both the coherent-spreading coefficient and the effective-attenuation coefficient. The net change in \(R_k\) is therefore determined by their competition. The canonical result is recovered when \(k_{\rm ph}\leq M\).

These results show how thermal occupation and modified mode frequencies affect Krylov-space spreading in the three inflationary models. The conclusions apply to the selected initial state, mode functions, parameter ranges, and effective coefficients. A microscopic system-environment interaction can be used in future work to derive the attenuation coefficient. Standard cosmological correlation functions can then be calculated within the same model.

\acknowledgments
LH is funded by the NSFC grant NO. 12665009 and Hunan Provincial Department of Education Project (Grant No.25B0480). TL is supported by the Guizhou Provincial Natural Science Foundation (Grant No.: 2025YJSKYJJ178) and the Natural Science Foundation of Guizhou Minzu University (Grant No.: GZMUBS7).

\bibliography{ref}

@article{Parker:2018yvk,
    author = "Parker, Daniel E. and Cao, Xiangyu and Avdoshkin, Alexander and Scaffidi, Thomas and Altman, Ehud",
    title = "{A Universal Operator Growth Hypothesis}",
    eprint = "1812.08657",
    archivePrefix = "arXiv",
    primaryClass = "cond-mat.stat-mech",
    doi = "10.1103/PhysRevX.9.041017",
    journal = "Phys. Rev. X",
    volume = "9",
    number = "4",
    pages = "041017",
    year = "2019"
}

@article{Armendariz-Picon:1999hyi,
    author = "Armendariz-Picon, C. and Damour, T. and Mukhanov, Viatcheslav F.",
    title = "{k - inflation}",
    eprint = "hep-th/9904075",
    archivePrefix = "arXiv",
    doi = "10.1016/S0370-2693(99)00603-6",
    journal = "Phys. Lett. B",
    volume = "458",
    pages = "209--218",
    year = "1999"
}

@article{Garriga:1999vw,
    author = "Garriga, Jaume and Mukhanov, Viatcheslav F.",
    title = "{Perturbations in k-inflation}",
    eprint = "hep-th/9904176",
    archivePrefix = "arXiv",
    reportNumber = "UAB-FT-466",
    doi = "10.1016/S0370-2693(99)00602-4",
    journal = "Phys. Lett. B",
    volume = "458",
    pages = "219--225",
    year = "1999"
}

@article{Bhattacharya:2023zqt,
    author = "Bhattacharya, Aranya and Nandy, Pratik and Nath, Pingal Pratyush and Sahu, Himanshu",
    title = "{On Krylov complexity in open systems: an approach via bi-Lanczos algorithm}",
    eprint = "2303.04175",
    archivePrefix = "arXiv",
    primaryClass = "quant-ph",
    reportNumber = "YITP-23-25",
    doi = "10.1007/JHEP12(2023)066",
    journal = "JHEP",
    volume = "12",
    pages = "066",
    year = "2023"
}

@article{Bhattacharya:2022gbz,
    author = "Bhattacharya, Aranya and Nandy, Pratik and Nath, Pingal Pratyush and Sahu, Himanshu",
    title = "{Operator growth and Krylov construction in dissipative open quantum systems}",
    eprint = "2207.05347",
    archivePrefix = "arXiv",
    primaryClass = "quant-ph",
    doi = "10.1007/JHEP12(2022)081",
    journal = "JHEP",
    volume = "12",
    pages = "081",
    year = "2022"
}

@article{Kofman:1994rk,
    author = "Kofman, Lev and Linde, Andrei D. and Starobinsky, Alexei A.",
    title = "{Reheating after inflation}",
    eprint = "hep-th/9405187",
    archivePrefix = "arXiv",
    reportNumber = "UH-IFA-94-35, SU-ITP-94-13, YITP-U-94-15",
    doi = "10.1103/PhysRevLett.73.3195",
    journal = "Phys. Rev. Lett.",
    volume = "73",
    pages = "3195--3198",
    year = "1994"
}

@article{Hikida:2004yc,
    author = "Hikida, Yasuaki",
    title = "{String theory on Lorentzian AdS(3) in minisuperspace}",
    eprint = "hep-th/0403081",
    archivePrefix = "arXiv",
    reportNumber = "SNUST-040301",
    doi = "10.1088/1126-6708/2004/04/025",
    journal = "JHEP",
    volume = "04",
    pages = "025",
    year = "2004"
}

@article{Naik:2001pr,
    author = "Naik, Satchidananda",
    title = "{The Sigma model formulation of the type 2 string theory in the Ads(5) x S(5) backgrounds with Ramond-Ramond flux}",
    eprint = "hep-th/0106116",
    archivePrefix = "arXiv",
    doi = "10.1142/S0217751X03014150",
    journal = "Int. J. Mod. Phys. A",
    volume = "18",
    pages = "1749--1760",
    year = "2003"
}

@article{Bojowald:2007cd,
    author = "Bojowald, Martin and Hossain, Golam Mortuza",
    title = "{Loop quantum gravity corrections to gravitational wave dispersion}",
    eprint = "0709.2365",
    archivePrefix = "arXiv",
    primaryClass = "gr-qc",
    reportNumber = "IGC-07-9-1",
    doi = "10.1103/PhysRevD.77.023508",
    journal = "Phys. Rev. D",
    volume = "77",
    pages = "023508",
    year = "2008"
}

@article{Barbon:2019wsy,
    author = "Barb{\'o}n, J. L. F. and Rabinovici, E. and Shir, R. and Sinha, R.",
    title = "{On The Evolution Of Operator Complexity Beyond Scrambling}",
    eprint = "1907.05393",
    archivePrefix = "arXiv",
    primaryClass = "hep-th",
    reportNumber = "IFT-UAM/CSIC-19-98",
    doi = "10.1007/JHEP10(2019)264",
    journal = "JHEP",
    volume = "10",
    pages = "264",
    year = "2019"
}

@article{Li:2024kfm,
    author = "Li, Tao and Liu, Lei-Hua",
    title = "{Inflationary Krylov complexity}",
    eprint = "2401.09307",
    archivePrefix = "arXiv",
    primaryClass = "hep-th",
    doi = "10.1007/JHEP04(2024)123",
    journal = "JHEP",
    volume = "04",
    pages = "123",
    year = "2024"
}

@article{Li:2023ekd,
    author = "Li, Tao and Liu, Lei-Hua",
    title = "{Cosmological complexity of the modified dispersion relation}",
    eprint = "2309.01595",
    archivePrefix = "arXiv",
    primaryClass = "gr-qc",
    doi = "10.1016/j.physletb.2024.138728",
    journal = "Phys. Lett. B",
    volume = "854",
    pages = "138728",
    year = "2024"
}

@article{Liu:2021nzx,
    author = "Liu, Lei-Hua and Li, Ai-Chen",
    title = "{Complexity of non-trivial sound speed in inflation}",
    eprint = "2102.12014",
    archivePrefix = "arXiv",
    primaryClass = "gr-qc",
    doi = "10.1016/j.dark.2022.101123",
    journal = "Phys. Dark Univ.",
    volume = "37",
    pages = "101123",
    year = "2022"
}

@article{Calcagni:2009ar,
    author = "Calcagni, Gianluca",
    title = "{Cosmology of the Lifshitz universe}",
    eprint = "0904.0829",
    archivePrefix = "arXiv",
    primaryClass = "hep-th",
    reportNumber = "IGC-09-4-2",
    doi = "10.1088/1126-6708/2009/09/112",
    journal = "JHEP",
    volume = "09",
    pages = "112",
    year = "2009"
}

@article{Kiritsis:2009sh,
    author = "Kiritsis, Elias and Kofinas, Georgios",
    title = "{Horava-Lifshitz Cosmology}",
    eprint = "0904.1334",
    archivePrefix = "arXiv",
    primaryClass = "hep-th",
    reportNumber = "CCTP-2009-16",
    doi = "10.1016/j.nuclphysb.2009.05.005",
    journal = "Nucl. Phys. B",
    volume = "821",
    pages = "467--480",
    year = "2009"
}

@article{Guth:1980zm,
    author = "Guth, Alan H.",
    editor = "Fang, Li-Zhi and Ruffini, R.",
    title = "{The Inflationary Universe: A Possible Solution to the Horizon and Flatness Problems}",
    reportNumber = "SLAC-PUB-2576",
    doi = "10.1103/PhysRevD.23.347",
    journal = "Phys. Rev. D",
    volume = "23",
    pages = "347--356",
    year = "1981"
}

@article{Cai:2009hc,
    author = "Cai, Yi-Fu and Zhang, Xinmin",
    title = "{Primordial perturbation with a modified dispersion relation}",
    eprint = "0906.3341",
    archivePrefix = "arXiv",
    primaryClass = "astro-ph.CO",
    doi = "10.1103/PhysRevD.80.043520",
    journal = "Phys. Rev. D",
    volume = "80",
    pages = "043520",
    year = "2009"
}

@article{Cai:2018tuh,
    author = "Cai, Yi-Fu and Tong, Xi and Wang, Dong-Gang and Yan, Sheng-Feng",
    title = "{Primordial Black Holes from Sound Speed Resonance during Inflation}",
    eprint = "1805.03639",
    archivePrefix = "arXiv",
    primaryClass = "astro-ph.CO",
    doi = "10.1103/PhysRevLett.121.081306",
    journal = "Phys. Rev. Lett.",
    volume = "121",
    number = "8",
    pages = "081306",
    year = "2018"
}

@article{Achucarro:2012sm,
    author = "Achucarro, Ana and Gong, Jinn-Ouk and Hardeman, Sjoerd and Palma, Gonzalo A. and Patil, Subodh P.",
    title = "{Effective theories of single field inflation when heavy fields matter}",
    eprint = "1201.6342",
    archivePrefix = "arXiv",
    primaryClass = "hep-th",
    reportNumber = "CERN-PH-TH-2011-222, CPHT-RR-055.0711, LPTENS-11-27",
    doi = "10.1007/JHEP05(2012)066",
    journal = "JHEP",
    volume = "05",
    pages = "066",
    year = "2012"
}

@article{Chen:2020uhe,
    author = "Chen, Chao and Ma, Xiao-Han and Cai, Yi-Fu",
    title = "{Dirac-Born-Infeld realization of sound speed resonance mechanism for primordial black holes}",
    eprint = "2003.03821",
    archivePrefix = "arXiv",
    primaryClass = "astro-ph.CO",
    doi = "10.1103/PhysRevD.102.063526",
    journal = "Phys. Rev. D",
    volume = "102",
    number = "6",
    pages = "063526",
    year = "2020"
}

@article{Susskind:2014moa,
    author = "Susskind, Leonard",
    title = "{Entanglement is not enough}",
    eprint = "1411.0690",
    archivePrefix = "arXiv",
    primaryClass = "hep-th",
    doi = "10.1002/prop.201500095",
    journal = "Fortsch. Phys.",
    volume = "64",
    pages = "49--71",
    year = "2016"
}

@article{Jefferson:2017sdb,
    author = "Jefferson, Ro and Myers, Robert C.",
    title = "{Circuit complexity in quantum field theory}",
    eprint = "1707.08570",
    archivePrefix = "arXiv",
    primaryClass = "hep-th",
    doi = "10.1007/JHEP10(2017)107",
    journal = "JHEP",
    volume = "10",
    pages = "107",
    year = "2017"
}

@article{Nielsen:2005mkt,
    author = "Nielsen, Michael A.",
    title = "{A geometric approach to quantum circuit lower bounds}",
    eprint = "quant-ph/0502070",
    archivePrefix = "arXiv",
    doi = "10.26421/QIC6.3-2",
    journal = "Quant. Inf. Comput.",
    volume = "6",
    number = "3",
    pages = "213--262",
    year = "2006"
}

@article{Hartman:2013qma,
    author = "Hartman, Thomas and Maldacena, Juan",
    title = "{Time Evolution of Entanglement Entropy from Black Hole Interiors}",
    eprint = "1303.1080",
    archivePrefix = "arXiv",
    primaryClass = "hep-th",
    doi = "10.1007/JHEP05(2013)014",
    journal = "JHEP",
    volume = "05",
    pages = "014",
    year = "2013"
}

@article{Susskind:2014rva,
    author = "Susskind, Leonard",
    title = "{Computational Complexity and Black Hole Horizons}",
    eprint = "1403.5695",
    archivePrefix = "arXiv",
    primaryClass = "hep-th",
    doi = "10.1002/prop.201500092",
    journal = "Fortsch. Phys.",
    volume = "64",
    pages = "24--43",
    year = "2016",
    note = "[Addendum: Fortsch.Phys. 64, 44--48 (2016)]"
}

@article{Zhao:2019nxk,
    author = "Zhao, Ying",
    title = "{A quantum circuit interpretation of evaporating black hole geometry}",
    eprint = "1912.00909",
    archivePrefix = "arXiv",
    primaryClass = "hep-th",
    doi = "10.1007/JHEP07(2020)139",
    journal = "JHEP",
    volume = "07",
    pages = "139",
    year = "2020"
}

@article{Chandra:2021kdv,
    author = {Chandra, A. Ramesh and de Boer, Jan and Flory, Mario and Heller, Michal P. and H{\"o}rtner, Sergio and Rolph, Andrew},
    title = "{Spacetime as a quantum circuit}",
    eprint = "2101.01185",
    archivePrefix = "arXiv",
    primaryClass = "hep-th",
    doi = "10.1007/JHEP04(2021)207",
    journal = "JHEP",
    volume = "21",
    pages = "207",
    year = "2021"
}

@article{Zhai:2025abc,
    author = "Zhai, Ke-Hong and Liu, Lei-Hua and Zhang, Hai-Qing",
    title = "{Inflationary power spectrum from the Lanczos algorithm}",
    eprint = "2505.20595",
    archivePrefix = "arXiv",
    primaryClass = "quant-ph",
    doi = "10.1140/epjc/s10052-025-14791-w",
    journal = "Eur. Phys. J. C",
    volume = "85",
    number = "10",
    pages = "1096",
    year = "2025"
}

@article{Liu:2026mzz,
    author = "Liu, Shi-Cheng and Liu, Lei-Hua and Li, Bichu and Zhang, Hai-Qing and He, Peng-Zhang",
    title = "{Quantum-information diagnostics of cosmological perturbations with nontrivial sound speed in inflation}",
    eprint = "2604.21755",
    archivePrefix = "arXiv",
    primaryClass = "gr-qc",
    month = "4",
    year = "2026"
}

@article{Ma:2026ksc,
    author = "Ma, Ming-Qi and Liu, Shi-Cheng and Liu, Lei-Hua and Zhang, Hai-Qing",
    title = "{Krylov complexity in non-inertial quantum systems}",
    eprint = "2606.29770",
    archivePrefix = "arXiv",
    primaryClass = "quant-ph",
    doi = "10.1140/epjc/s10052-026-16271-1",
    journal = "Eur. Phys. J. C",
    volume = "86",
    number = "9",
    pages = "1092",
    year = "2026"
}

@article{Liu:2026cxb,
    author = "Liu, Shi-Cheng and Liu, Lei-Hua and Li, Bichu and Zhou, Yuebing and Zhang, Hai-Qing",
    title = "{A Bogoliubov-ratio framework for quantum-information diagnostics of time-dependent two-mode Boson System}",
    eprint = "2607.24404",
    archivePrefix = "arXiv",
    primaryClass = "quant-ph",
    month = "7",
    year = "2026"
}

@article{Zheng:2026vne,
    author = "Zheng, Han-Qi and He, Peng-Zhang and Liu, Lei-Hua and Zhang, Hai-Qing",
    title = "{Geometric-phase control of Krylov complexity in adiabatic dynamics}",
    eprint = "2608.14325",
    archivePrefix = "arXiv",
    primaryClass = "quant-ph",
     journal = "",
    month = "8",
    year = "2026"
}

@article{Jafarizadeh:2006woc,
    author = "Jafarizadeh, M. A. and Salimi, S. and Sufiani, R.",
    title = "{Investigation of continuous-time quantum walk by using Krylov subspace-Lanczos algorithm}",
    eprint = "quant-ph/0606241",
    archivePrefix = "arXiv",
    doi = "10.1140/epjb/e2007-00281-5",
     journal = "",
    month = "6",
    year = "2006"
}

@article{Li:2024ljz,
    author = "Li, Tao and Liu, Lei-Hua",
    title = "{Krylov complexity of thermal state in early universe}",
    eprint = "2408.03293",
    archivePrefix = "arXiv",
    primaryClass = "hep-th",
    doi = "10.1140/epjc/s10052-026-15490-w",
    journal = "Eur. Phys. J. C",
    volume = "86",
    number = "3",
    pages = "265",
    year = "2026"
}

@article{Zhai:2024tkz,
    author = "Zhai, Ke-Hong and Liu, Lei-Hua and Zhang, Hai-Qing",
    title = "{Generalized CV Conjecture and Krylov Complexity in Two-Mode Hermitian Systems via Information Geometry}",
    eprint = "2412.08925",
    archivePrefix = "arXiv",
    primaryClass = "hep-th",
    doi = "10.1016/j.aop.2026.170534",
    journal = "Annals Phys.",
    volume = "491",
    pages = "170534",
    year = "2026"
}

@article{Kar:2021nbm,
    author = "Kar, Arjun and Lamprou, Lampros and Rozali, Moshe and Sully, James",
    title = "{Random matrix theory for complexity growth and black hole interiors}",
    eprint = "2106.02046",
    archivePrefix = "arXiv",
    primaryClass = "hep-th",
    doi = "10.1007/JHEP01(2022)016",
    journal = "JHEP",
    volume = "01",
    pages = "016",
    year = "2022"
}

@article{Liu:2025caj,
    author = "Liu, Shi-Cheng and Liu, Lei-Hua and Li, Bichu and Zhang, Hai-Qing and He, Peng-Zhang",
    title = "{A quantum information method for early universe with non-trivial sound speed}",
    eprint = "2510.04011",
    archivePrefix = "arXiv",
    primaryClass = "gr-qc",
    doi = "10.1002/prop.70081",
    journal = "Fortsch. Phys.",
    volume = "74",
    pages = "e70081",
    year = "2026"
}

@article{Muck:2022xfc,
    author = {M{\"u}ck, Wolfgang and Yang, Yi},
    title = "{Krylov complexity and orthogonal polynomials}",
    eprint = "2205.12815",
    archivePrefix = "arXiv",
    primaryClass = "hep-th",
    doi = "10.1016/j.nuclphysb.2022.115948",
    journal = "Nucl. Phys. B",
    volume = "984",
    pages = "115948",
    year = "2022"
}

@article{Rabinovici:2020ryf,
    author = "Rabinovici, E. and S{\'a}nchez-Garrido, A. and Shir, R. and Sonner, J.",
    title = "{Operator complexity: a journey to the edge of Krylov space}",
    eprint = "2009.01862",
    archivePrefix = "arXiv",
    primaryClass = "hep-th",
    doi = "10.1007/JHEP06(2021)062",
    journal = "JHEP",
    volume = "06",
    pages = "062",
    year = "2021"
}

@article{Li:2026jxx,
    author = "Li, Tao and Fu, Hai-Bing",
    title = "{Circuit and Krylov complexity of primordial perturbations of modified gravity in inflation}",
    eprint = "2607.09408",
    archivePrefix = "arXiv",
    primaryClass = "hep-th",
    journal = "",
    month = "7",
    year = "2026"
}

@article{Adhikari:2022oxr,
    author = "Adhikari, Kiran and Choudhury, Sayantan",
    title = "{Cosmological Krylov Complexity}",
    eprint = "2203.14330",
    archivePrefix = "arXiv",
    primaryClass = "hep-th",
    doi = "10.1002/prop.202200126",
    journal = "Fortsch. Phys.",
    volume = "70",
    number = "12",
    pages = "2200126",
    year = "2022"
}

@article{Chakraborty:2023lpr,
    author = "Chakraborty, Ayan and Maity, Debaprasad",
    title = "{Squeezing, chaos and thermalization in periodically driven quantum systems: the case of bosonic preheating}",
    eprint = "2309.10116",
    archivePrefix = "arXiv",
    primaryClass = "hep-th",
    doi = "10.1007/JHEP02(2024)110",
    journal = "JHEP",
    volume = "02",
    pages = "110",
    year = "2024"
}

@article{Maldacena:2015waa,
    author = "Maldacena, Juan and Shenker, Stephen H. and Stanford, Douglas",
    title = "{A bound on chaos}",
    eprint = "1503.01409",
    archivePrefix = "arXiv",
    primaryClass = "hep-th",
    doi = "10.1007/JHEP08(2016)106",
    journal = "JHEP",
    volume = "08",
    pages = "106",
    year = "2016"
}

@article{Linardopoulos:2022wol,
    author = "Linardopoulos, Georgios",
    title = "{String integrability of the ABJM defect}",
    eprint = "2202.06824",
    archivePrefix = "arXiv",
    primaryClass = "hep-th",
    doi = "10.1007/JHEP06(2022)033",
    journal = "JHEP",
    volume = "06",
    pages = "033",
    year = "2022"
}

@article{Berera:1995ie,
    author = "Berera, Arjun",
    title = "{Warm inflation}",
    eprint = "astro-ph/9509049",
    archivePrefix = "arXiv",
    reportNumber = "PSU-TH-159",
    doi = "10.1103/PhysRevLett.75.3218",
    journal = "Phys. Rev. Lett.",
    volume = "75",
    pages = "3218--3221",
    year = "1995"
}

@article{Jian:2020qpp,
    author = "Jian, Shao-Kai and Swingle, Brian and Xian, Zhuo-Yu",
    title = "{Complexity growth of operators in the SYK model and in JT gravity}",
    eprint = "2008.12274",
    archivePrefix = "arXiv",
    primaryClass = "hep-th",
    doi = "10.1007/JHEP03(2021)014",
    journal = "JHEP",
    volume = "03",
    pages = "014",
    year = "2021"
}

@article{He:2022ryk,
    author = "He, Song and Lau, Pak Hang Chris and Xian, Zhuo-Yu and Zhao, Long",
    title = "{Quantum chaos, scrambling and operator growth in $ T\overline{T} $ deformed SYK models}",
    eprint = "2209.14936",
    archivePrefix = "arXiv",
    primaryClass = "hep-th",
    doi = "10.1007/JHEP12(2022)070",
    journal = "JHEP",
    volume = "12",
    pages = "070",
    year = "2022"
}

@article{Basu:2025ubf,
    author = "Basu, Pallab and Das, Suman and Nandy, Pratik",
    title = "{Complexity of quadratic quantum chaos}",
    eprint = "2509.04075",
    archivePrefix = "arXiv",
    primaryClass = "hep-th",
    reportNumber = "RIKEN-iTHEMS-Report-25",
    doi = "10.1007/JHEP04(2026)081",
    journal = "JHEP",
    volume = "04",
    pages = "081",
    year = "2026"
}

@article{Erdmenger:2023wjg,
    author = "Erdmenger, Johanna and Jian, Shao-Kai and Xian, Zhuo-Yu",
    title = "{Universal chaotic dynamics from Krylov space}",
    eprint = "2303.12151",
    archivePrefix = "arXiv",
    primaryClass = "hep-th",
    doi = "10.1007/JHEP08(2023)176",
    journal = "JHEP",
    volume = "08",
    pages = "176",
    year = "2023"
}

@article{Patramanis:2023cwz,
    author = "Patramanis, Dimitrios and Sybesma, Watse",
    title = {{Krylov complexity in a natural basis for the Schr{\"o}dinger algebra}},
    eprint = "2306.03133",
    archivePrefix = "arXiv",
    primaryClass = "quant-ph",
    doi = "10.21468/SciPostPhysCore.7.2.037",
    journal = "SciPost Phys. Core",
    volume = "7",
    pages = "037",
    year = "2024"
}

@article{Hashimoto:2023swv,
    author = "Hashimoto, Koji and Murata, Keiju and Tanahashi, Norihiro and Watanabe, Ryota",
    title = "{Krylov complexity and chaos in quantum mechanics}",
    eprint = "2305.16669",
    archivePrefix = "arXiv",
    primaryClass = "hep-th",
    reportNumber = "KUNS-2967",
    doi = "10.1007/JHEP11(2023)040",
    journal = "JHEP",
    volume = "11",
    pages = "040",
    year = "2023"
}

@article{Vasli:2023syq,
    author = "Vasli, M. J. and Babaei Velni, K. and Mohammadi Mozaffar, M. R. and Mollabashi, A. and Alishahiha, M.",
    title = "{Krylov complexity in Lifshitz-type scalar field theories}",
    eprint = "2307.08307",
    archivePrefix = "arXiv",
    primaryClass = "hep-th",
    reportNumber = "IPM/P-2023/53",
    doi = "10.1140/epjc/s10052-024-12609-9",
    journal = "Eur. Phys. J. C",
    volume = "84",
    number = "3",
    pages = "235",
    year = "2024"
}

@article{Domingo:2023kjr,
    author = "Domingo, Laia and Borondo, F. and Scialchi, Gast{\'o}n and Roncaglia, Augusto J. and Carlo, Gabriel G. and Wisniacki, Diego A.",
    title = "{Quantum reservoir complexity by the Krylov evolution approach}",
    eprint = "2310.00790",
    archivePrefix = "arXiv",
    primaryClass = "quant-ph",
    doi = "10.1103/PhysRevA.110.022446",
    journal = "Phys. Rev. A",
    volume = "110",
    number = "2",
    pages = "022446",
    year = "2024"
}

@article{Gill:2023umm,
    author = "Gill, Ankit and Pal, Kunal and Pal, Kuntal and Sarkar, Tapobrata",
    title = "{Complexity in two-point measurement schemes}",
    eprint = "2311.07892",
    archivePrefix = "arXiv",
    primaryClass = "quant-ph",
    doi = "10.1103/PhysRevB.109.104303",
    journal = "Phys. Rev. B",
    volume = "109",
    number = "10",
    pages = "104303",
    year = "2024"
}

@article{Bhattacharjee:2023uwx,
    author = "Bhattacharjee, Budhaditya and Nandy, Pratik and Pathak, Tanay",
    title = "{Operator dynamics in Lindbladian SYK: a Krylov complexity perspective}",
    eprint = "2311.00753",
    archivePrefix = "arXiv",
    primaryClass = "quant-ph",
    reportNumber = "YITP-23-133, RIKEN-iTHEMS-Report-23",
    doi = "10.1007/JHEP01(2024)094",
    journal = "JHEP",
    volume = "01",
    pages = "094",
    year = "2024"
}

@article{Adhikari:2022whf,
    author = "Adhikari, Kiran and Choudhury, Sayantan and Roy, Abhishek",
    title = "{Krylov Complexity in Quantum Field Theory}",
    eprint = "2204.02250",
    archivePrefix = "arXiv",
    primaryClass = "hep-th",
    doi = "10.1016/j.nuclphysb.2023.116263",
    journal = "Nucl. Phys. B",
    volume = "993",
    pages = "116263",
    year = "2023"
}

@article{Camargo:2023eev,
    author = "Camargo, Hugo A. and Jahnke, Viktor and Jeong, Hyun-Sik and Kim, Keun-Young and Nishida, Mitsuhiro",
    title = "{Spectral and Krylov complexity in billiard systems}",
    eprint = "2306.11632",
    archivePrefix = "arXiv",
    primaryClass = "hep-th",
    reportNumber = "IFT-UAM/CSIC-23-61",
    doi = "10.1103/PhysRevD.109.046017",
    journal = "Phys. Rev. D",
    volume = "109",
    number = "4",
    pages = "046017",
    year = "2024"
}

@article{Bhattacharjee:2022vlt,
    author = "Bhattacharjee, Budhaditya and Cao, Xiangyu and Nandy, Pratik and Pathak, Tanay",
    title = "{Krylov complexity in saddle-dominated scrambling}",
    eprint = "2203.03534",
    archivePrefix = "arXiv",
    primaryClass = "quant-ph",
    doi = "10.1007/JHEP05(2022)174",
    journal = "JHEP",
    volume = "05",
    pages = "174",
    year = "2022"
}

@article{Huh:2023jxt,
    author = "Huh, Kyoung-Bum and Jeong, Hyun-Sik and Pedraza, Juan F.",
    title = "{Spread complexity in saddle-dominated scrambling}",
    eprint = "2312.12593",
    archivePrefix = "arXiv",
    primaryClass = "hep-th",
    reportNumber = "IFT-UAM/CSIC-23-178",
    doi = "10.1007/JHEP05(2024)137",
    journal = "JHEP",
    volume = "05",
    pages = "137",
    year = "2024"
}

@article{Li:2021kfq,
    author = "Li, Ai-chen and Li, Xin-Fei and Zeng, Ding-fang and Liu, Lei-Hua",
    title = "{Cosmological complexity in K-essence}",
    eprint = "2102.12939",
    archivePrefix = "arXiv",
    primaryClass = "gr-qc",
    doi = "10.1016/j.dark.2024.101422",
    journal = "Phys. Dark Univ.",
    volume = "43",
    pages = "101422",
    year = "2024"
}

@article{Armendariz-Picon:2003jjq,
    author = "Armendariz-Picon, Christian and Lim, Eugene A.",
    title = "{Scale invariance without inflation?}",
    eprint = "astro-ph/0307101",
    archivePrefix = "arXiv",
    doi = "10.1088/1475-7516/2003/12/002",
    journal = "JCAP",
    volume = "12",
    pages = "002",
    year = "2003"
}

@article{Armendariz-Picon:2006vgx,
    author = "Armendariz-Picon, Cristian",
    title = "{Near Scale Invariance with Modified Dispersion Relations}",
    eprint = "astro-ph/0606168",
    archivePrefix = "arXiv",
    doi = "10.1088/1475-7516/2006/10/010",
    journal = "JCAP",
    volume = "10",
    pages = "010",
    year = "2006"
}

@article{Magueijo:2008sx,
    author = "Magueijo, Joao",
    title = "{Bimetric varying speed of light theories and primordial fluctuations}",
    eprint = "0807.1689",
    archivePrefix = "arXiv",
    primaryClass = "gr-qc",
    doi = "10.1103/PhysRevD.79.043525",
    journal = "Phys. Rev. D",
    volume = "79",
    pages = "043525",
    year = "2009"
}

@article{Martin:2000xs,
    author = "Martin, Jerome and Brandenberger, Robert H.",
    title = "{The TransPlanckian problem of inflationary cosmology}",
    eprint = "hep-th/0005209",
    archivePrefix = "arXiv",
    doi = "10.1103/PhysRevD.63.123501",
    journal = "Phys. Rev. D",
    volume = "63",
    pages = "123501",
    year = "2001"
}

@article{Arkani-Hamed:2003pdi,
    author = "Arkani-Hamed, Nima and Cheng, Hsin-Chia and Luty, Markus A. and Mukohyama, Shinji",
    title = "{Ghost condensation and a consistent infrared modification of gravity}",
    eprint = "hep-th/0312099",
    archivePrefix = "arXiv",
    reportNumber = "HUTP-03-A081, UMD-PPP-04-012",
    doi = "10.1088/1126-6708/2004/05/074",
    journal = "JHEP",
    volume = "05",
    pages = "074",
    year = "2004"
}

@article{Bojowald:2006zb,
    author = "Bojowald, Martin and Hernandez, Hector and Kagan, Mikhail and Singh, Parampreet and Skirzewski, Aureliano",
    title = "{Formation and Evolution of Structure in Loop Cosmology}",
    eprint = "astro-ph/0611685",
    archivePrefix = "arXiv",
    reportNumber = "IGPG-06-11-3, AEI-2006-085",
    doi = "10.1103/PhysRevLett.98.031301",
    journal = "Phys. Rev. Lett.",
    volume = "98",
    pages = "031301",
    year = "2007"
}

@article{Jacobson:2000gw,
    author = "Jacobson, Ted and Mattingly, David",
    title = "{Generally covariant model of a scalar field with high frequency dispersion and the cosmological horizon problem}",
    eprint = "hep-th/0009052",
    archivePrefix = "arXiv",
    doi = "10.1103/PhysRevD.63.041502",
    journal = "Phys. Rev. D",
    volume = "63",
    pages = "041502",
    year = "2001"
}

@article{Cai:2007gs,
    author = "Cai, Yi-fu and Li, Ming-zhe and Lu, Jian-Xin and Piao, Yun-Song and Qiu, Tao-tao and Zhang, Xin-min",
    title = "{A String-Inspired Quintom Model Of Dark Energy}",
    eprint = "hep-th/0701016",
    archivePrefix = "arXiv",
    reportNumber = "USTC-ICTS-07-03",
    doi = "10.1016/j.physletb.2007.05.056",
    journal = "Phys. Lett. B",
    volume = "651",
    pages = "1--7",
    year = "2007"
}

@article{Cai:2009in,
    author = "Cai, Yi-Fu and Saridakis, Emmanuel N.",
    title = "{Non-singular cosmology in a model of non-relativistic gravity}",
    eprint = "0906.1789",
    archivePrefix = "arXiv",
    primaryClass = "hep-th",
    doi = "10.1088/1475-7516/2009/10/020",
    journal = "JCAP",
    volume = "10",
    pages = "020",
    year = "2009"
}

@article{Li:2009rt,
    author = "Li, Mingzhe and Cai, Yi-Fu and Wang, Xiulian and Zhang, Xinmin",
    title = "{$CPT$ Violating Electrodynamics and Chern-Simons Modified Gravity}",
    eprint = "0907.5159",
    archivePrefix = "arXiv",
    primaryClass = "hep-ph",
    doi = "10.1016/j.physletb.2009.08.053",
    journal = "Phys. Lett. B",
    volume = "680",
    pages = "118--124",
    year = "2009"
}

@article{Cai:2009zp,
    author = "Cai, Yi-Fu and Saridakis, Emmanuel N. and Setare, Mohammad R. and Xia, Jun-Qing",
    title = "{Quintom Cosmology: Theoretical implications and observations}",
    eprint = "0909.2776",
    archivePrefix = "arXiv",
    primaryClass = "hep-th",
    doi = "10.1016/j.physrep.2010.04.001",
    journal = "Phys. Rept.",
    volume = "493",
    pages = "1--60",
    year = "2010"
}

@article{Cai:2012yf,
    author = "Cai, Yi-Fu and Li, Mingzhe and Zhang, Xinmin",
    title = "{Emergent Universe Scenario via Quintom Matter}",
    eprint = "1209.3437",
    archivePrefix = "arXiv",
    primaryClass = "hep-th",
    doi = "10.1016/j.physletb.2012.10.065",
    journal = "Phys. Lett. B",
    volume = "718",
    pages = "248--254",
    year = "2012"
}

@article{Zheng:2017qfs,
    author = "Zheng, Yunlong and Shen, Liuyuan and Mou, Yicen and Li, Mingzhe",
    title = "{On (in)stabilities of perturbations in mimetic models with higher derivatives}",
    eprint = "1704.06834",
    archivePrefix = "arXiv",
    primaryClass = "gr-qc",
    reportNumber = "USTC-ICTS-17-04",
    doi = "10.1088/1475-7516/2017/08/040",
    journal = "JCAP",
    volume = "08",
    pages = "040",
    year = "2017"
}

@article{Chen:2017dhi,
    author = "Chen, Jun and Hou, Wenjie and Hou, Defu and Qiu, Taotao",
    title = "{Comparing potential-driven DBI-inspired non-minimal kinetic coupling (Dinkic) inflation with observational data}",
    eprint = "1711.06580",
    archivePrefix = "arXiv",
    primaryClass = "astro-ph.CO",
    doi = "10.1088/1674-1137/42/4/045102",
    journal = "Chin. Phys. C",
    volume = "42",
    number = "4",
    pages = "045102",
    year = "2018"
}

@article{Bianco:2016yib,
    author = "Bianco, Stefano and Friedhoff, Victor Nicolai and Wilson-Ewing, Edward",
    title = "{Modified dispersion relations, inflation and scale invariance}",
    eprint = "1609.06891",
    archivePrefix = "arXiv",
    primaryClass = "gr-qc",
    doi = "10.1103/PhysRevD.97.046006",
    journal = "Phys. Rev. D",
    volume = "97",
    number = "4",
    pages = "046006",
    year = "2018"
}

@article{Pan:2015tza,
    author = "Pan, Wen-Jian and Huang, Yong-Chang",
    title = "{Bouncing universe with modified dispersion relation}",
    eprint = "1508.06475",
    archivePrefix = "arXiv",
    primaryClass = "hep-th",
    doi = "10.1007/s10714-016-2138-y",
    journal = "Gen. Rel. Grav.",
    volume = "48",
    number = "11",
    pages = "144",
    year = "2016"
}

@article{Patramanis:2021lkx,
    author = "Patramanis, Dimitrios",
    title = "{Probing the entanglement of operator growth}",
    eprint = "2111.03424",
    archivePrefix = "arXiv",
    primaryClass = "hep-th",
    doi = "10.1093/ptep/ptac081",
    journal = "PTEP",
    volume = "2022",
    number = "6",
    pages = "063A01",
    year = "2022"
}

@article{Cao:2020zls,
    author = "Cao, Xiangyu",
    title = "{A statistical mechanism for operator growth}",
    eprint = "2012.06544",
    archivePrefix = "arXiv",
    primaryClass = "cond-mat.stat-mech",
    doi = "10.1088/1751-8121/abe77c",
    journal = "J. Phys. A",
    volume = "54",
    number = "14",
    pages = "144001",
    year = "2021"
}

@article{Trigueros:2021rwj,
    author = "Trigueros, Fabian Ballar and Lin, Cheng-Ju",
    title = "{Krylov complexity of many-body localization: Operator localization in Krylov basis}",
    eprint = "2112.04722",
    archivePrefix = "arXiv",
    primaryClass = "cond-mat.dis-nn",
    doi = "10.21468/SciPostPhys.13.2.037",
    journal = "SciPost Phys.",
    volume = "13",
    number = "2",
    pages = "037",
    year = "2022"
}

@article{Heveling:2022hth,
    author = "Heveling, Robin and Wang, Jiaozi and Gemmer, Jochen",
    title = "{Numerically probing the universal operator growth hypothesis}",
    eprint = "2203.00533",
    archivePrefix = "arXiv",
    primaryClass = "cond-mat.stat-mech",
    doi = "10.1103/PhysRevE.106.014152",
    journal = "Phys. Rev. E",
    volume = "106",
    number = "1",
    pages = "014152",
    year = "2022"
}

@article{Dymarsky:2021bjq,
    author = "Dymarsky, Anatoly and Smolkin, Michael",
    title = "{Krylov complexity in conformal field theory}",
    eprint = "2104.09514",
    archivePrefix = "arXiv",
    primaryClass = "hep-th",
    doi = "10.1103/PhysRevD.104.L081702",
    journal = "Phys. Rev. D",
    volume = "104",
    number = "8",
    pages = "L081702",
    year = "2021"
}

@article{Caputa:2021ori,
    author = "Caputa, Pawel and Datta, Shouvik",
    title = "{Operator growth in 2d CFT}",
    reportNumber = "CERN-TH-2021-151",
    doi = "10.1007/JHEP12(2021)188",
    journal = "JHEP",
    volume = "12",
    pages = "188",
    year = "2021",
    note = "[Erratum: JHEP 09, 113 (2022)]"
}

@article{Caputa:2022eye,
    author = "Caputa, Pawel and Liu, Sinong",
    title = "{Quantum complexity and topological phases of matter}",
    eprint = "2205.05688",
    archivePrefix = "arXiv",
    primaryClass = "hep-th",
    doi = "10.1103/PhysRevB.106.195125",
    journal = "Phys. Rev. B",
    volume = "106",
    number = "19",
    pages = "195125",
    year = "2022"
}

@article{He:2025guu,
    author = "He, Peng-Zhang and Liu, Lei-Hua and Zhang, Hai-Qing and Jiang, Qing-Quan",
    title = {{Krylov complexity and Wightman power spectrum with positive chemical potential in Schr{\"o}dinger field theory}},
    eprint = "2509.14742",
    archivePrefix = "arXiv",
    primaryClass = "hep-th",
    doi = "10.1007/JHEP02(2026)259",
    journal = "JHEP",
    volume = "02",
    pages = "259",
    year = "2026"
}

@article{Ali:2018fcz,
    author = "Ali, Tibra and Bhattacharyya, Arpan and Shajidul Haque, S. and Kim, Eugene H. and Moynihan, Nathan",
    title = "{Time Evolution of Complexity: A Critique of Three Methods}",
    eprint = "1810.02734",
    archivePrefix = "arXiv",
    primaryClass = "hep-th",
    reportNumber = "YITP-18-111",
    doi = "10.1007/JHEP04(2019)087",
    journal = "JHEP",
    volume = "04",
    pages = "087",
    year = "2019"
}

@article{Haque:2021kdm,
    author = "Haque, S. Shajidul and Jana, Chandan and Underwood, Bret",
    title = "{Saturation of thermal complexity of purification}",
    eprint = "2107.08969",
    archivePrefix = "arXiv",
    primaryClass = "hep-th",
    doi = "10.1007/JHEP01(2022)159",
    journal = "JHEP",
    volume = "01",
    pages = "159",
    year = "2022"
}

@article{Das:2024zuu,
    author = "Das, Rathindra Nath and Mori, Takato",
    title = "{Krylov Complexity of Purification}",
    eprint = "2408.00826",
    archivePrefix = "arXiv",
    primaryClass = "hep-th",
    reportNumber = "YITP-24-94",
    doi = "10.1103/qgcx-wxpd",
    journal = "Phys. Rev. Lett.",
    volume = "136",
    number = "3",
    pages = "030201",
    year = "2026"
}

\end{document}